\documentclass{article} 
\usepackage{iclr2025_conference,times}
\iclrfinalcopy

\usepackage{amsmath,amsfonts,bm}

\def\eqref#1{equation~\ref{#1}}

\def\1{\bm{1}}

\DeclareMathAlphabet{\mathsfit}{\encodingdefault}{\sfdefault}{m}{sl}
\SetMathAlphabet{\mathsfit}{bold}{\encodingdefault}{\sfdefault}{bx}{n}

\usepackage{hyperref}
\usepackage{url}

\usepackage{microtype}
\usepackage{hyperref}
\usepackage{url}
\usepackage{booktabs}
\usepackage{makecell}
\usepackage{graphicx}
\usepackage{multicol}
\usepackage{multirow}
\usepackage{tabularx}
\usepackage{etoc}
\usepackage{subfigure}
\usepackage{wrapfig}
\usepackage{lineno}

\title{A Generalist Audio Foundation Model for Comprehensive Body Sound Auscultation}

\author{
Pingjie Wang\textsuperscript{1,3}, Liudan Zhao\textsuperscript{2}, Zihan Zhao\textsuperscript{1,3}, Miao He\textsuperscript{4}, Xin Sun\textsuperscript{2}, \textbf{Ya Zhang\textsuperscript{1,3}}  \\
\textbf{Kun Sun\textsuperscript{2}, Yanfeng Wang\textsuperscript{1,3}}, \textbf{Yu Wang\textsuperscript{1,3}\thanks{Corresponding Author.}} \\
\textsuperscript{1} Shanghai Jiao Tong University \\
\textsuperscript{2} Xinhua Hospital Affiliated To Shanghai Jiao Tong University School of Medicine \\
\textsuperscript{3} Shanghai Artificial Intelligence Laboratory \\
\textsuperscript{4} Tongji University \\
\texttt{\{pingjiewang, yuwangsjtu\}@sjtu.edu.com} \\
}

\begin{document}

\etocdepthtag.toc{mtchapter}
\etocsettagdepth{mtchapter}{subsection}
\etocsettagdepth{mtappendix}{none}

\maketitle
\begin{abstract}
Accurate and efficient auscultation-based diagnostics are vital for early disease detection, especially in resource-limited settings where specialized clinical expertise is scarce. Traditional auscultation, which heavily depends on clinician experience, suffers from significant inter-observer variability, while existing AI models often falter due to the limitations of non-representative training data. In this study, we introduce AuscultaBase, a novel AI-driven diagnostic framework that harnesses self-supervised and contrastive learning techniques alongside large-scale, multi-source data integration to advance body sound analysis. By generating robust feature representations, AuscultaBase markedly enhances performance in abnormality detection, disease classification, and activity recognition tasks. Comprehensive evaluations on our newly established benchmark, AuscultaBench, demonstrate that AuscultaBase consistently outperforms state-of-the-art methods across key performance metrics, underscoring its potential as a scalable and cost-effective tool for clinical screening and early disease intervention. The code and model checkpoint has been released in \url{https://github.com/applewpj/AuscultaBase}.


\end{abstract}

\section{Introduction}

Auscultation, the clinical practice of listening to internal body sounds, is a cornerstone of non-invasive diagnosis for a range of health conditions. By analyzing heart, respiratory, and bowel sounds, clinicians gain critical insights into a patient’s physiological state, enabling the detection and monitoring of disorders such as valvular heart disease, congenital heart defects, pneumonia, asthma, gastroenteritis, and intestinal obstruction. However, the accuracy and consistency of auscultation remain heavily reliant on clinician expertise and the limited acoustic sensitivity of the human ear, which can result in variability and missed diagnoses \citep{mangione1997cardiac}. The growing need for standardized, efficient, and widely accessible diagnostic tools has driven research into artificial intelligence (AI)-based auscultation analysis, which offers the potential to enhance diagnostic accuracy and scalability \citep{chen2023robust,kong2020panns,oletic2017asthmatic,zhang2024towards,sitaula2022neonatal}.

Despite recent progress, several fundamental challenges remain unaddressed. Existing methods predominantly rely on general acoustic techniques that are not optimized for auscultatory signals, limiting their ability to capture the pathological variations in body sounds \citep{chen2023robust,guo2023ds,oletic2017asthmatic,jung2021efficiently}. While unsupervised neural network training has demonstrated effectiveness in learning high-quality audio representations \citep{baevski2020wav2vec,huang2022masked,hsu2021hubert,wu2023large}, pre-trained audio models are often developed using datasets such as AudioSet \citep{gemmeke2017audio} and LibriSpeech \citep{panayotov2015librispeech}, which primarily contain music and speech rather than medically relevant auscultatory sounds. This domain gap diminishes their applicability for body sound analysis. Furthermore, publicly available datasets currently span only a fraction of the diversity observed in clinical auscultation, further constraining the generalization ability of AI-driven models.

Given the critical need for rapid and cost-effective screening solutions, especially in resource-limited settings, we propose a comprehensive framework designed to advance AI-assisted auscultation. Our approach aims to enhance the diagnostic capabilities of primary care providers while enabling scalable, high-quality body sound analysis. The framework is structured around three core components:
\begin{itemize}
\item  \textbf{AuscultaCorpus}: A large-scale, multi-source body sound database. This corpus constitutes an extensive repository of body sound recordings, integrating ten publicly available datasets with one proprietary dataset. This amalgamation encompasses a diverse spectrum of auscultatory sounds, including cardiac, respiratory, and gastrointestinal recordings, totaling 322.4 hours of training data. The corpus offers a comprehensive and representative foundation for the development and evaluation of diagnostic models across various auscultation systems, thereby facilitating advancements in automated diagnostic methodologies.
\item \textbf{AuscultaBase}: A foundational diagnostic  model for body sounds. AuscultaBase is developed as a robust diagnostic framework specifically designed for body sound analysis. Built on contrastive learning techniques applied to the AuscultaCorpus, the model effectively captures intricate acoustic patterns inherent in physiological sounds. Comparative evaluations demonstrate significant enhancements in diagnostic accuracy relative to traditional acoustic models. This unified approach enables the simultaneous analysis of diverse body sounds, thereby establishing a new benchmark for precision and reliability in auscultation-based diagnostics.
\item \textbf{AuscultaBench}: A comprehensive evaluation benchmark to assess AI-based auscultation models. To facilitate rigorous assessment, AuscultaBench provides a structured evaluation framework across two primary diagnostic tasks—abnormality detection and disease classification—spanning 16 sub-tasks that reflect the diverse and complex demands of clinical auscultation. This benchmark ensures consistent and reproducible performance comparisons, facilitating advancements in automated diagnostic methodologies.
\end{itemize}

\begin{figure}[t]
    \centering
    \includegraphics[width=\linewidth]{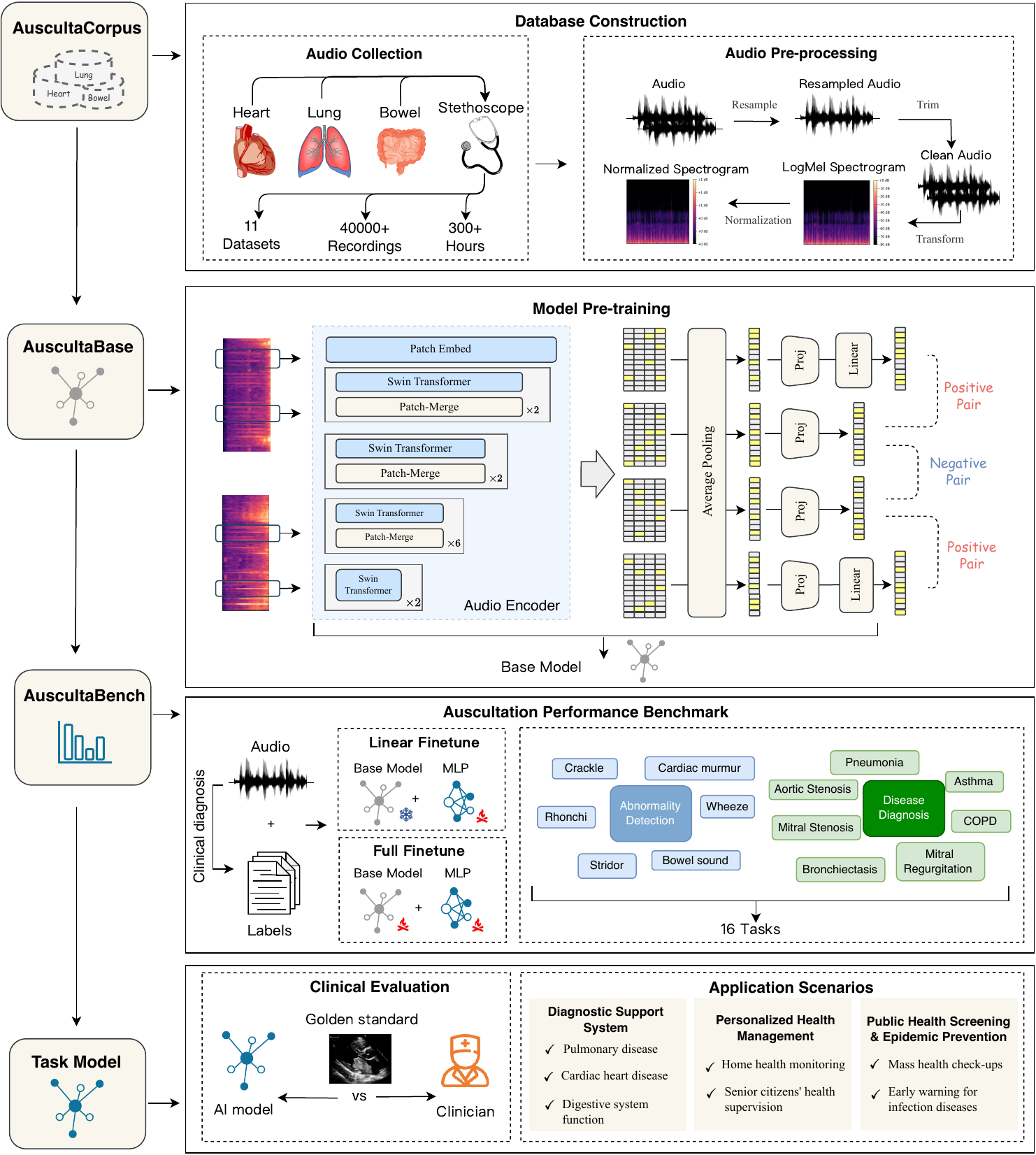}
    \caption{System overview of our framework. During the Data Construction stage, the AuscultaCorpus is first collected by stethoscopes and then pre-processed into spectrograms. Followed by the AuscultaCorpus, the audio encoder is pre-trained with contrastive learning to obtain the base model for auscultation, AuscultaBase. Finally, we apply linear/full fine-tuning to AuscultaBase to adapt to 16 downstream auscultation tasks within our evaluation benchmark, AuscultaBench. Additionally, we conduct clinical evaluations between our model and experienced clinicians to analyze the auscultation performance, extending the prospect for various scenarios of clinical applications.}
    \label{fig:overview}
\end{figure}

\section{Results}

\subsection{Overview of proposed datasets \& model \& benchmark}
The system overview of our framework is illustrated in Figure~\ref{fig:overview}, comprising three key components: the pretraining dataset (AuscultaCorpus), the foundation model (AuscultaBase), and the evaluation benchmark (AuscultaBench). AuscultaCorpus is constructed by aggregating diverse body sound datasets and preprocessing them for subsequent model pretraining.


AuscultaCorpus is curated from 11 datasets to enable the pretraining of our auscultation model, covering lung, heart, and bowel sounds. After filtering out corrupted audio clips, the final pretraining dataset comprises 40,317 samples with a total duration of 322.4 hours. The statistics of these datasets are summarized in Extended Table~\ref{tab:pretraining_datasets}.


We pretrain our auscultation model using the constructed datasets to build our auscultation foundation model: AuscultaBase. For datasets that already include distinct training and validation splits, we preserve the provided validation sets. For those without an explicit division, we randomly select 10\% of the samples for validation after shuffling the data.

\begin{figure}[t]
    \centering
    \includegraphics[width=1.1\linewidth]{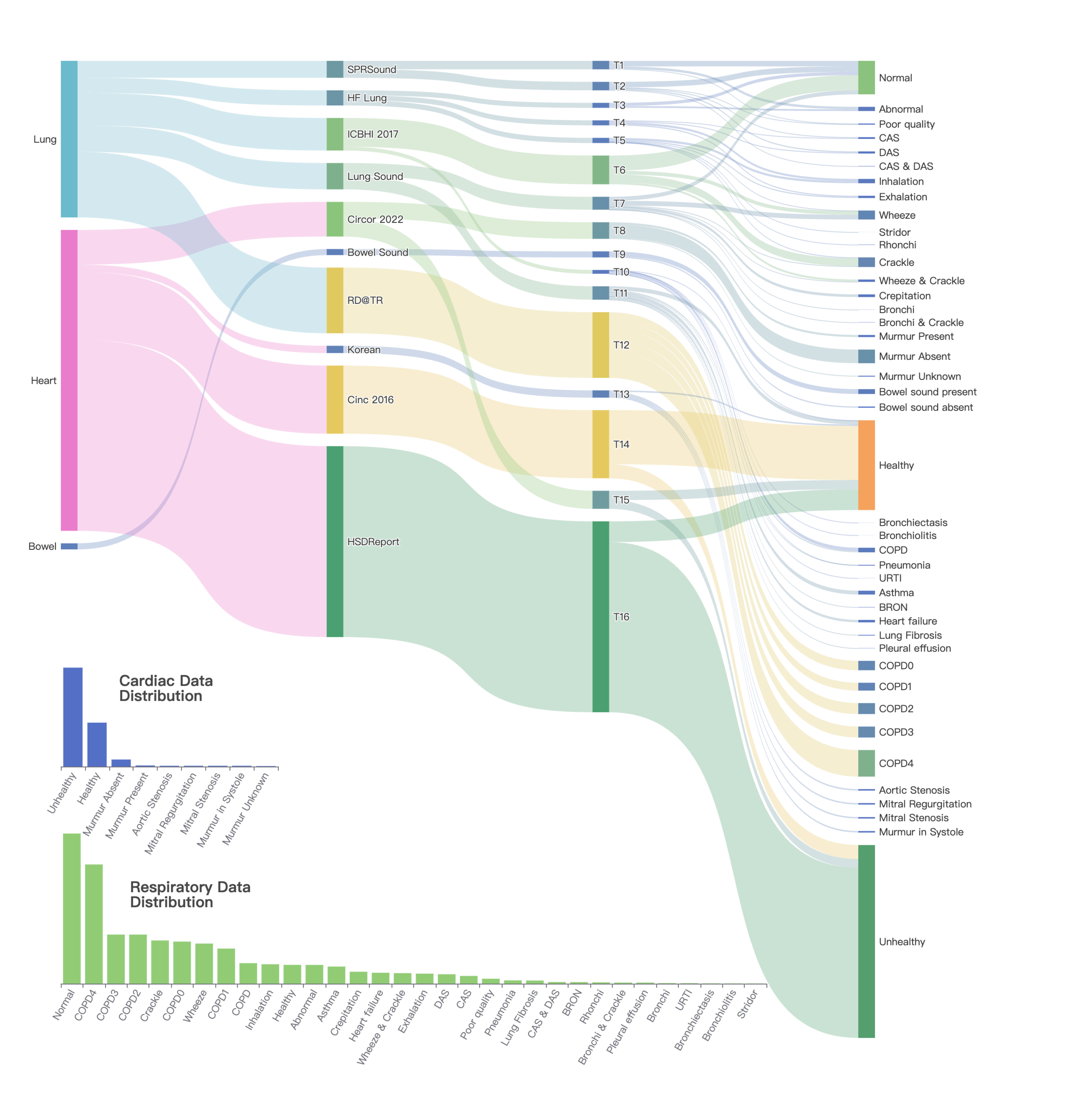}
    \caption{Statistics of AuscultaBench. The Sankey diagram shows how the different body sound types (left), datasets (left middle), tasks (right middle), and abnormality or disease categories (right) contribute to the final evaluation benchmark. On the left of the bottom, two bar charts show the data distributions on cardiac and respiratory entities respectively.}
    \label{fig:benchmark_sanky}
\end{figure}

To evaluate the auscultation capabilities of current foundation models, we introduce AuscultaBench—a comprehensive benchmark derived from the datasets collated. The statistics of AuscultaBench is exhibited in Figure~\ref{fig:benchmark_sanky}, and a more specific distribution is detailed in Extended Table~\ref{tab:pretraining_datasets}. Importantly, all test sets originally associated with these datasets were excluded beforehand to prevent any data leakage.


\subsection{Overall Performance Across Diverse Tasks}

Tables~\ref{tab:overall_performance} summarize the Macro-F1 and Micro-F1, Precision, and Recall scores, respectively. Our proposed model, AuscultaBase, demonstrates clear and consistent improvements over the baseline models across these varied tasks. Notably, AuscultaBase achieved the highest Macro-F1 performance in 12 out of 16 tasks and led in Micro-F1 for 9 tasks. By comparison, the baselines—such as OPERA-CT, AudioMAE, CLAP, and PANN—led in at most a single task each. For instance, PANN led only in tasks T6 and T10 for Micro-F1, and even then, its performance did not translate to improvements in Macro-F1. This consistent superiority across different body sound categories (lung, heart, and bowel) underscores the enhanced capability of AuscultaBase to extract distinctive and generalizable acoustic features, which are critical for accurate and reliable body sound classification.

\paragraph{Balanced Accuracy Across Imbalanced Datasets}
A further comparison between Macro-F1 and Micro-F1 scores reveals that AuscultaBase maintains balanced accuracy across classes, an essential quality given the inherent class imbalances in many medical datasets. In 9 out of 16 tasks (T1, T3, T7, T9, T11–T16), AuscultaBase simultaneously achieved the highest scores for both metrics. This balanced performance is particularly significant in clinical contexts, where ensuring that minority classes (often representing critical or rare conditions) are accurately identified is as important as the overall performance.

\begin{table}[htbp]
    \centering
    \resizebox{\linewidth}{0.75\linewidth}{
        \begin{tabular}{lcccccccc}
            \toprule
            \textbf{Task ID} & \textbf{Function} & \textbf{\makecell[c]{Sound\\Type}} & \textbf{\makecell[c]{Task\\Type}} & \textbf{OPERA-CT} & \textbf{AudioMAE} & \textbf{CLAP} & \textbf{PANN} & \textbf{AuscultaBase} \\
    
            \midrule
            \midrule
            \multicolumn{9}{c}{\textit{Macro-F1}} \\
            \midrule
            T1 & AD & L & MC & 44.99$\pm$1.53 & 38.14$\pm$1.24 & 47.71$\pm$0.09 & 25.99$\pm$0.07 & \textbf{48.14}$\pm$0.46 \\
            T2 & AD & L & MC & 25.09$\pm$1.04 & 18.20$\pm$0.81 & \textbf{36.55}$\pm$0.41 & 15.55$\pm$0.00 & 30.25$\pm$1.10 \\
            T3 & AD & L & BC & 54.49$\pm$1.36 & 55.72$\pm$0.11 & 55.63$\pm$0.63 & 40.99$\pm$0.00 & \textbf{57.28}$\pm$0.78 \\
            T4 & AD & L & MC & 32.08$\pm$0.05 & 27.70$\pm$0.14 & 33.19$\pm$0.03 & 16.75$\pm$0.03 & \textbf{33.48}$\pm$0.12 \\
            T5 & AD & L & MC & 22.04$\pm$0.08 & 14.42$\pm$0.00 & 23.24$\pm$0.65 & 11.29$\pm$0.15 & \textbf{23.74}$\pm$0.01 \\
            T6 & AD  & L & MC & 28.65$\pm$0.87 & 19.99$\pm$1.33 & \textbf{37.90}$\pm$0.26 & 34.71$\pm$0.31 & 26.23$\pm$0.35 \\
            T7 & AD  & L & MC & 23.95$\pm$0.13 & 9.34$\pm$1.53 & 27.53$\pm$2.91 & 7.80$\pm$0.00 & \textbf{28.03}$\pm$3.66 \\
            T8 & AD  & H & MC & 44.40$\pm$0.52 & 42.93$\pm$2.16 & \textbf{48.33}$\pm$3.18 & 28.71$\pm$0.00 & 41.52$\pm$2.45 \\
            T9 & AD  & B & BC & 69.82$\pm$1.38 & 45.60$\pm$0.85 & 71.99$\pm$1.02 & 66.40$\pm$1.29 & \textbf{72.88}$\pm$0.05	 \\
            T10 & DD  & L & MC & 23.28$\pm$4.08 & 22.10$\pm$2.97 & \textbf{34.37}$\pm$3.83 & 21.46$\pm$2.03 & 24.68$\pm$0.83 \\
            T11 & DD & L & ML & 15.10$\pm$7.12 & 7.31$\pm$4.18 & 7.99$\pm$0.07 & 0.00$\pm$0.00 & \textbf{20.10}$\pm$5.85 \\
            T12 & DD & L & MC & 30.74$\pm$0.77 & 17.00$\pm$5.17 & 27.68$\pm$2.07 & 30.15$\pm$5.57 & \textbf{41.78}$\pm$1.83 \\
            T13 & DD & H & MC & 77.37$\pm$1.94 & 50.18$\pm$15.23 & 86.82$\pm$0.32 & 84.19$\pm$2.15 & \textbf{88.10}$\pm$0.69 \\
            T14 & DD & H & BC & 65.04$\pm$0.56 & 72.35$\pm$0.62 & 61.97$\pm$2.78 & 66.01$\pm$0.34 & \textbf{77.84}$\pm$0.23 \\
            T15 & DD & H & BC & 58.10$\pm$1.14 & 61.66$\pm$0.45 & 59.94$\pm$1.12 & 48.38$\pm$0.37 & \textbf{62.84}$\pm$1.49 \\
            T16 & DD & H & BC & 47.27$\pm$0.00 & 47.27$\pm$0.00 & 47.27$\pm$0.00 & 47.27$\pm$0.00 & \textbf{52.53}$\pm$1.15 \\

            \midrule
            \midrule
            \multicolumn{9}{c}{\textit{Micro-F1}} \\
            \midrule
             T1 & AD  & L & MC & 64.24$\pm$1.81 & 61.68$\pm$4.55 & \textbf{72.09}$\pm$1.68 & 63.64$\pm$0.03 & 71.28$\pm$0.37 \\
            T2 & AD  & L & MC & 64.02$\pm$0.22 & 63.21$\pm$0.16 & 61.03$\pm$0.97 & 63.61$\pm$0.00 & \textbf{65.73}$\pm$0.12 \\
            T3 & AD  & L & BC & \textbf{71.24}$\pm$0.34 & 63.58$\pm$0.44 & 66.22$\pm$1.34 & 69.45$\pm$0.00 & \textbf{70.67}$\pm$0.17 \\
            T4 & AD  & L & MC & \textbf{51.32}$\pm$0.29 & 46.54$\pm$0.19 & 44.84$\pm$0.35 & 49.60$\pm$0.01 & 50.46$\pm$0.21 \\
            T5 & AD  & L & MC & \textbf{51.03}$\pm$0.31 & 48.45$\pm$0.00 & 44.50$\pm$0.54 & 49.59$\pm$0.02 & 49.42$\pm$1.07 \\
            T6 & AD  & L & MC & 46.77$\pm$2.54 & 32.98$\pm$5.55 & 53.86$\pm$0.38 & \textbf{54.86}$\pm$0.15 & 41.75$\pm$0.82 \\
            T7 & AD  & L & MC & 48.48$\pm$0.00 & 25.76$\pm$1.52 & 46.97$\pm$1.52 & 24.24$\pm$0.00 & \textbf{51.52}$\pm$3.03 \\
            T8 & AD  & H & MC & 77.53$\pm$0.00 & 80.06$\pm$0.32 & \textbf{81.33}$\pm$0.32 & 75.63$\pm$0.00 & 78.16$\pm$0.32 \\
            T9 & AD  & B & BC & 82.24$\pm$0.93 & 81.15$\pm$0.16 & 82.87$\pm$0.62 & 82.55$\pm$0.31 & \textbf{85.05}$\pm$0.31	 \\
            T10 & DD & L & MC & 90.81$\pm$0.26 & 91.34$\pm$0.26 & 91.73$\pm$0.66 & \textbf{92.13}$\pm$0.00 & 90.81$\pm$1.31 \\
            T11 & DD & L & ML & 18.49$\pm$8.78 & 22.52$\pm$8.88 & 19.22$\pm$0.30 & 0.00$\pm$0.00 & \textbf{43.15}$\pm$1.29 \\
            T12 & DD & L & MC & 36.00$\pm$2.00 & 43.00$\pm$1.00 & 41.00$\pm$3.00 & 40.00$\pm$4.00 & \textbf{49.00}$\pm$1.00 \\
            T13 & DD & H & MC & 77.00$\pm$2.00 & 56.50$\pm$9.50 & \textbf{87.50}$\pm$0.50 & 84.00$\pm$2.00 & \textbf{87.50}$\pm$0.50 \\
            T14 & DD & H & BC & 81.83$\pm$0.47 & 81.52$\pm$0.47 & 80.90$\pm$0.16 & 76.86$\pm$0.31 & \textbf{87.27}$\pm$0.31 \\
            T15 & DD & H & BC & 59.02$\pm$2.06 & 62.03$\pm$0.63 & 60.92$\pm$1.11 & 55.70$\pm$0.32 & \textbf{63.45}$\pm$1.42 \\
            T16 & DD & H & BC & 89.66$\pm$0.00 & 89.66$\pm$0.00 & 89.66$\pm$0.00 & 89.66$\pm$0.00 & \textbf{90.52}$\pm$0.65 \\ 
            
            \midrule
            \midrule
            \multicolumn{9}{c}{\textit{Precision}} \\
            \midrule
            T1 & AD  & L & MC & 50.80$\pm$2.60 & 48.00$\pm$7.06 & 48.80$\pm$1.01 & 29.55$\pm$8.35 & \textbf{50.86}$\pm$0.31 \\
            T2 & AD  & L & MC & \textbf{39.39}$\pm$8.63 & 24.73$\pm$0.18 & 38.52$\pm$1.75 & 12.72$\pm$0.00 & 38.77$\pm$0.73  \\
            T3 & AD  & L & BC & \textbf{65.93}$\pm$1.51 & 56.00$\pm$0.06 & 57.57$\pm$0.68 & 34.73$\pm$0.00 & 63.92$\pm$0.41 \\
            T4 & AD  & L & MC & 40.11$\pm$0.39 & 32.97$\pm$0.35 & 35.55$\pm$0.18 & 18.60$\pm$0.07 & \textbf{42.11}$\pm$0.08 \\
            T5 & AD  & L & MC & 33.04$\pm$0.93 & 26.00$\pm$0.00 & 28.77$\pm$0.36 & 12.68$\pm$0.24 & \textbf{33.33}$\pm$0.14 \\
            T6 & AD  & L & MC & 32.36$\pm$5.23 & 24.93$\pm$2.36 & \textbf{38.42}$\pm$0.20 & 36.38$\pm$2.44 & 30.31$\pm$0.13 \\
            T7 & AD  & L & MC & 19.84$\pm$0.16 & 10.00$\pm$5.16 & \textbf{38.13}$\pm$0.87 & 4.85$\pm$0.00 & 28.42$\pm$6.25 \\
            T8 & AD  & H & MC & 64.57$\pm$0.44 & 54.66$\pm$2.81 & \textbf{71.40}$\pm$17.25 & 25.21$\pm$0.00 & 56.75$\pm$1.96 \\
            T9 & AD  & B & BC & 70.88$\pm$1.63 & 65.56$\pm$25.06 & 72.14$\pm$1.02 & 71.27$\pm$0.71 & \textbf{76.50}$\pm$0.92 \\
            T10 & DD & L & MC & 22.81$\pm$4.26 & 21.50$\pm$3.18 & \textbf{45.01}$\pm$1.54 & 32.64$\pm$5.97 & 26.16$\pm$0.73 \\
            T11 & DD & L & ML & 12.56$\pm$7.64 & 5.41$\pm$2.00 & 10.91$\pm$0.00 & 0.00$\pm$0.00 & \textbf{30.00}$\pm$7.50  \\
            T12 & DD & L & MC & \textbf{40.90}$\pm$3.90 & 13.57$\pm$5.18 & 33.68$\pm$1.26 & 34.71$\pm$11.37 & 38.78$\pm$0.28 \\
            T13 & DD & H & MC & 77.59$\pm$1.51 & 58.90$\pm$15.83 & 87.38$\pm$0.43 & 84.08$\pm$2.34 & \textbf{89.65}$\pm$1.04 \\
            T14 & DD & H & BC & 74.70$\pm$2.18 & 72.47$\pm$0.50 & 72.60$\pm$0.79 & 65.66$\pm$0.42 & \textbf{84.55}$\pm$1.22 \\
            T15 & DD & H & BC & 59.42$\pm$2.17 & 61.90$\pm$0.66 & 60.91$\pm$1.19 & 57.21$\pm$0.61 & \textbf{63.41}$\pm$1.46 \\
            T16 & DD & H & BC & 44.83$\pm$0.00 & 44.83$\pm$0.00 & 44.83$\pm$0.00 & 44.83$\pm$0.00 & \textbf{78.41}$\pm$16.61 \\
    
            \midrule
            \midrule
            \multicolumn{9}{c}{\textit{Recall}} \\
            \midrule
            T1 & AD  & L & MC & 44.12$\pm$1.24 & 39.71$\pm$1.33 & 48.31$\pm$0.23 & 33.36$\pm$0.03 & \textbf{47.40}$\pm$0.52 \\
            T2 & AD  & L & MC & 25.82$\pm$0.82 & 21.10$\pm$0.52 & \textbf{38.89}$\pm$0.47 & 20.00$\pm$0.00 & 29.03$\pm$0.85 \\
            T3 & AD  & L & BC & 56.13$\pm$1.35 & 55.62$\pm$0.13 & 55.74$\pm$0.35 & 50.00$\pm$0.00 & \textbf{57.63}$\pm$0.48 \\
            T4 & AD  & L & MC & 32.80$\pm$0.13 & 30.06$\pm$0.84 & \textbf{33.74}$\pm$0.00 & 25.05$\pm$0.00 & 33.43$\pm$0.14 \\
            T5 & AD  & L & MC & 22.60$\pm$0.10 & 17.71$\pm$0.00 & \textbf{23.85}$\pm$0.73 & 16.75$\pm$0.05 & 23.29$\pm$0.17  \\
            T6 & AD  & L & MC & 28.89$\pm$0.32 & 25.75$\pm$2.61 & \textbf{38.89}$\pm$0.03 & 35.47$\pm$0.02 & 28.47$\pm$0.23 \\
            T7 & AD  & L & MC & 30.87$\pm$0.48 & 20.77$\pm$0.77 & 30.09$\pm$3.31 & 20.00$\pm$0.00 & \textbf{34.05}$\pm$1.74 \\
            T8 & AD  & H & MC & 42.02$\pm$0.36 & 41.91$\pm$1.75 & \textbf{45.32}$\pm$1.74 & 33.33$\pm$0.00 & 40.52$\pm$1.64 \\
            T9 & AD  & B & BC & 68.96$\pm$1.20 & 50.41$\pm$0.41 & \textbf{71.86}$\pm$1.01 & 64.23$\pm$1.35 & 70.69$\pm$0.44 \\
            T10 & DD & L & MC & 25.92$\pm$6.03 & 22.74$\pm$2.74 & \textbf{30.01}$\pm$3.78 & 20.49$\pm$1.86 & 24.97$\pm$0.75 \\
            T11 & DD & L & ML & \textbf{28.22}$\pm$0.72 & 20.19$\pm$17.31 & 8.59$\pm$2.27 & 0.00$\pm$0.00 & 16.55$\pm$5.29 \\
            T12 & DD & L & MC & 43.17$\pm$0.32 & 25.00$\pm$5.00 & 31.27$\pm$2.14 & 33.85$\pm$4.72 & \textbf{52.62}$\pm$1.67 \\
            T13 & DD & H & MC & 79.72$\pm$1.66 & 56.68$\pm$14.33 & 86.86$\pm$0.36 & 84.70$\pm$1.69 & \textbf{87.48}$\pm$0.59 \\
            T14 & DD & H & BC & 62.88$\pm$0.54 & 72.74$\pm$1.25 & 60.43$\pm$2.25 & 66.73$\pm$0.54 & \textbf{74.43}$\pm$0.07 \\
            T15 & DD & H & BC & 58.64$\pm$1.42 & 61.68$\pm$0.49 & 60.22$\pm$1.10 & 53.73$\pm$0.32 & \textbf{62.92}$\pm$1.46 \\
            T16 & DD & H & BC & 50.00$\pm$0.00 & 50.00$\pm$0.00 & 50.00$\pm$0.00 & 50.00$\pm$0.00 & \textbf{52.64}$\pm$0.56 \\

            \bottomrule
        \end{tabular}
    }
    \caption{Macro-F1, Micro-F1, Precision, and Recall scores for abnormality detection (AD: T1-T8) and disease diagnosis (DD: T10-T16) tasks. The best model for each task is in \textbf{bold}. We report mean and standard deviation with 3 runs. Macro-F1 ranges from 0 to 100, and \textcolor{red}{higher} is better.}
    
    \label{tab:overall_performance}
\end{table}

\subsection{Binary Classification: Discriminative Power and AUC Analysis}

\begin{figure}[t]
    \centering
    \includegraphics[width=\linewidth]{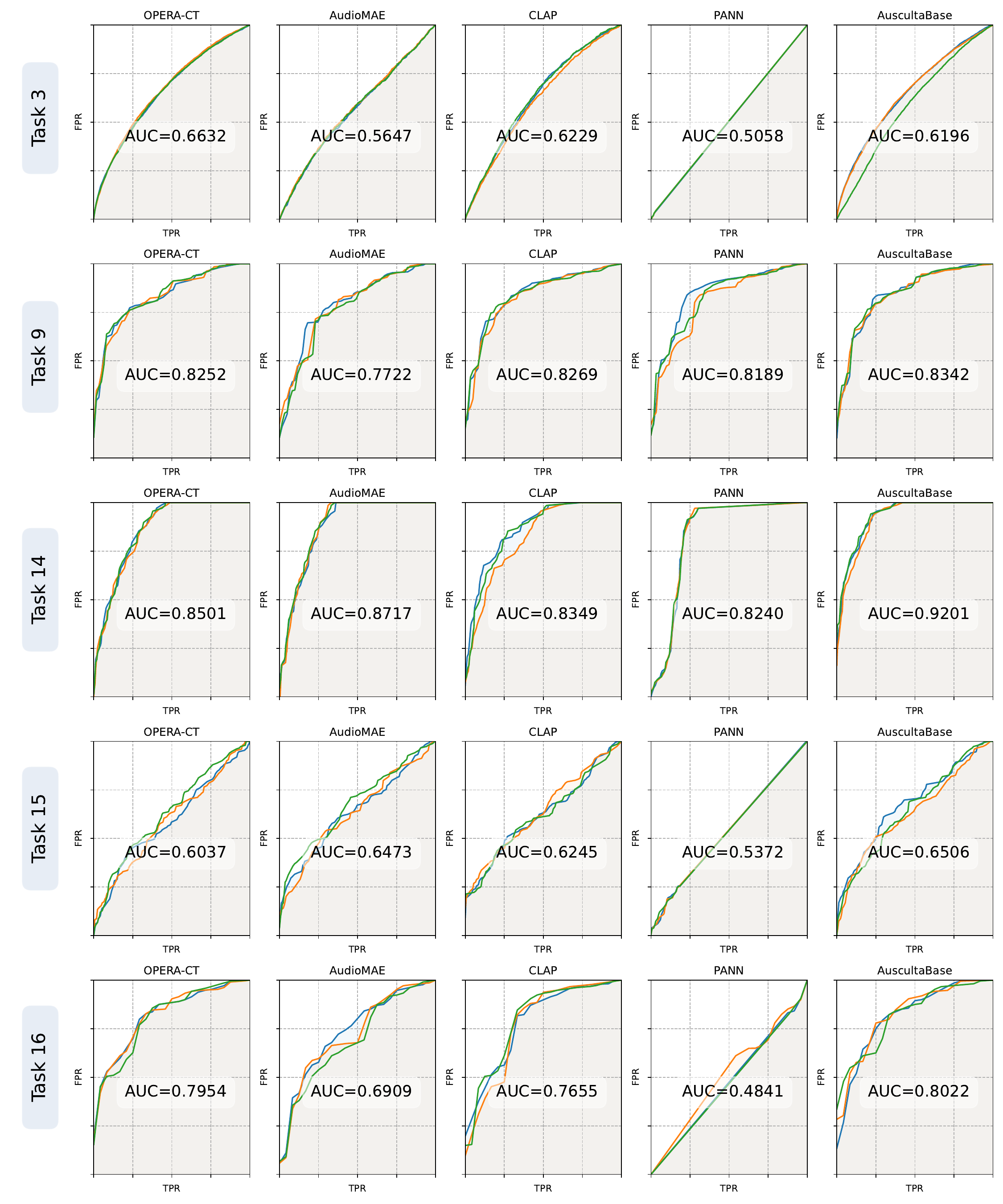}
    \caption{Receiver operating characteristic (ROC) curve and average AUC of binary classification tasks (T4, T9, T14, T15, and T16) with 3 independent runs (labeled with 3 colors).}
    \label{fig:AUC}
\end{figure}

To further probe the model’s discriminative capabilities, we computed the Area Under the Receiver Operating Characteristic curve (AUC) for select binary classification tasks (T3, T9, T14, T15, and T16). Unlike F1 scores, AUC provides an aggregate measure of performance across all classification thresholds. As shown in Figure~\ref{fig:AUC}, AuscultaBase consistently outperformed all baseline models, particularly in task T14, where it achieved an AUC of 0.920 compared to 0.872 for AudioMAE (the best baseline). This superior discriminative performance is critical for clinical applications, as it indicates a robust ability to distinguish between healthy and abnormal states, thus reducing the risk of misdiagnosis.

\begin{figure}[htbp]
    \centering
    \includegraphics[width=\linewidth]{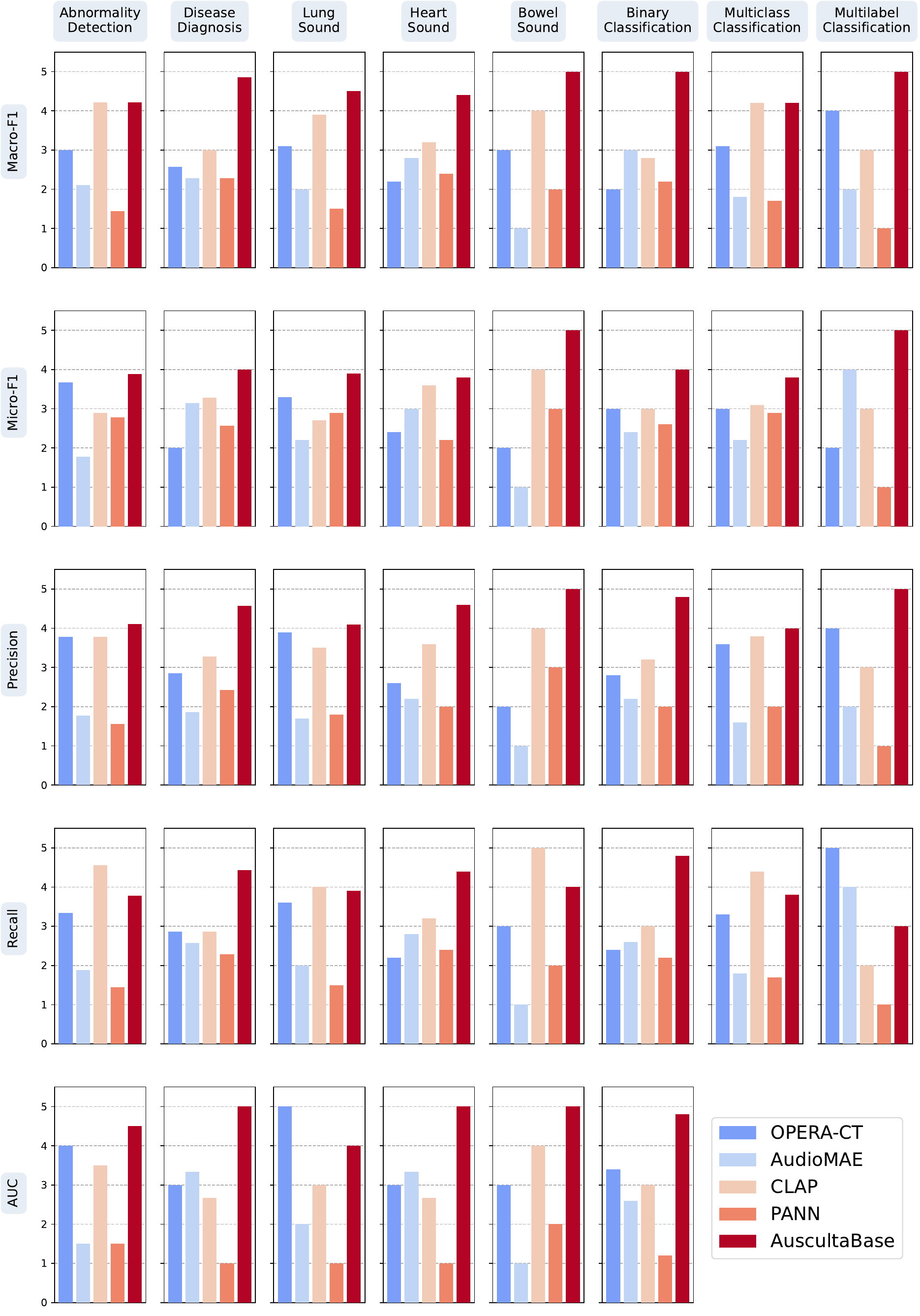}
    \caption{Borda count scores (BCS) categorized by the task function (abnormality detection and disease diagnosis), sound type (lung sound, heart sound, and bowel sound), and task type (binary classification, multiclass classification, and multilabel classification). BCS ranges from 1 to 5, and \textcolor{red}{higher} is better.}
    \label{fig:categorized_bcs}
\end{figure}

\subsection{Multidimensional Evaluation of Model Robustness}

To provide a nuanced evaluation of model performance across multiple dimensions, we categorized the tasks by function, sound type, and task type, and computed the Borda Count Scores (BCS) based on the reported Macro-F1, Micro-F1, Precision, Recall, and AUC performance as depicted in Figure~\ref{fig:categorized_bcs}.

\paragraph{Performance by Function: Abnormality Detection and Disease Diagnosis}
When tasks were grouped into abnormality detection and disease diagnosis, AuscultaBase always achieved the highest BCS in F1 scores and AUC. In disease diagnosis tasks, for example, AuscultaBase reaches a BCS of 4.86 (Macro-F1), a substantial improvement over competing models. This result not only highlights the model’s capacity for nuanced clinical assessment but also its potential to impact patient care through more precise disease identification.

\paragraph{Adaptability Across Sound Types}
Performance was also evaluated based on the sound type, using BCS for lung, heart, and bowel sounds. As depicted in Figure~\ref{fig:categorized_bcs}, AuscultaBase consistently outperformed baseline models, particularly excelling in heart sound classification. This robust performance across diverse acoustic environments demonstrates the model’s adaptability, making it a versatile tool for real-world clinical applications where multiple sound sources and conditions are encountered.

\paragraph{Robustness Across Task Types}
Finally, we assessed performance across various task formulations, including binary classification, multi-class classification, and multi-label classification. As shown in the right panel of Figure~\ref{fig:categorized_bcs}, AuscultaBase achieved top performance across all task types, with particularly strong results in binary classification tasks. This versatility indicates that the model is capable of effectively generalizing to different problem formulations, an important quality for its application in diverse diagnostic scenarios.

\subsection{Comparative Analysis: AuscultaBase vs. Human Experts}

\subsubsection{Clinical Evaluation Setup}

\begin{figure}[t]
    \centering
    \subfigure[]{
        \includegraphics[width=0.95\linewidth]{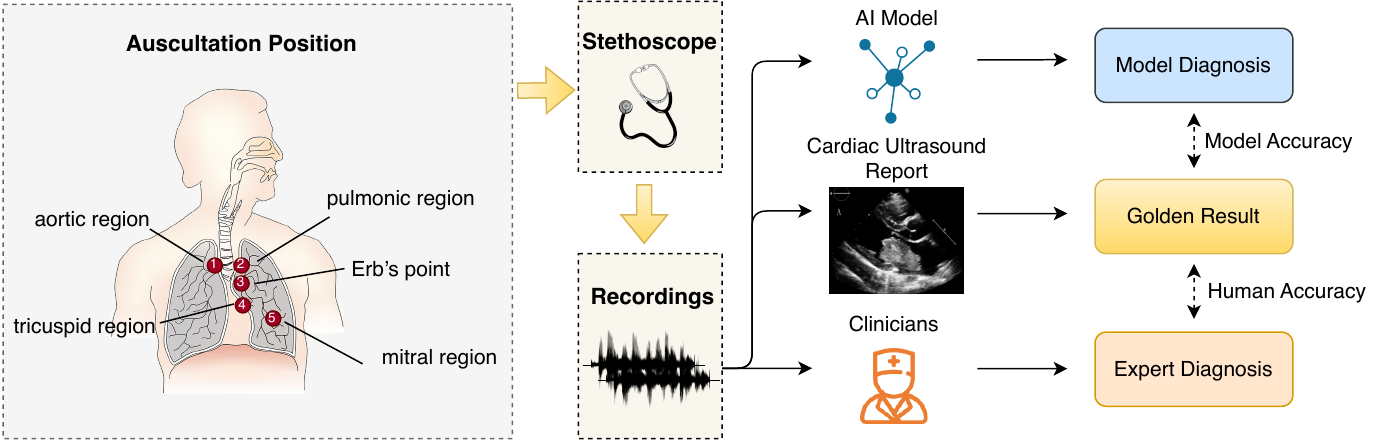}
        \label{fig:human_eval_process}
    }
    \subfigure[]{
        \includegraphics[width=0.21\linewidth]{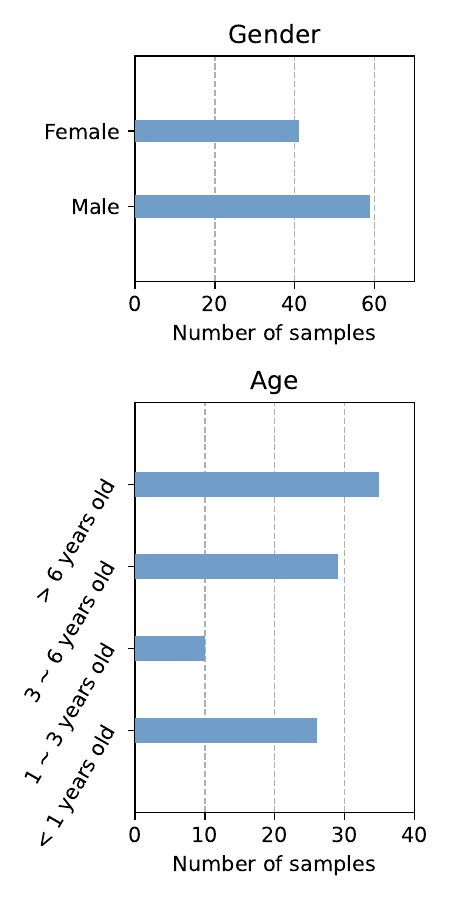}
        \label{fig:human_eval_statistics}
    }\hspace{-10pt}
    \subfigure[]{
        \includegraphics[width=0.21\linewidth]{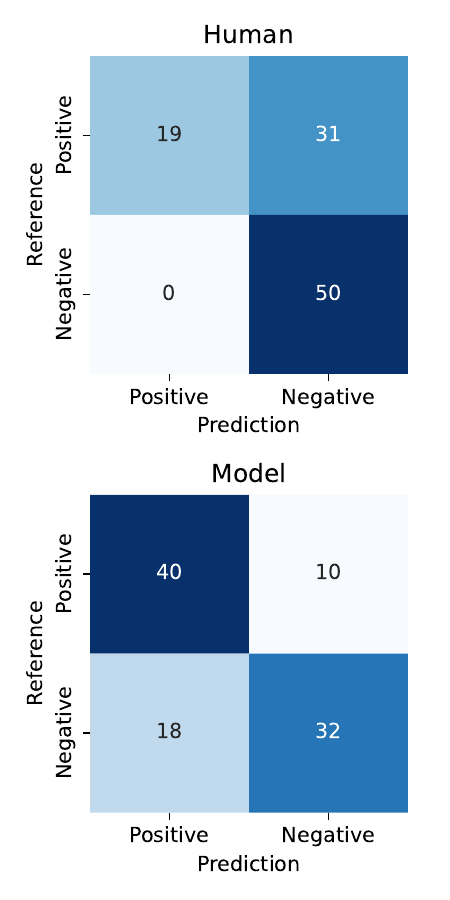}
        \label{fig:human_eval_confusion_matrix}
    }\hspace{-10pt}
    \subfigure[]{
        \includegraphics[width=0.5775\linewidth]{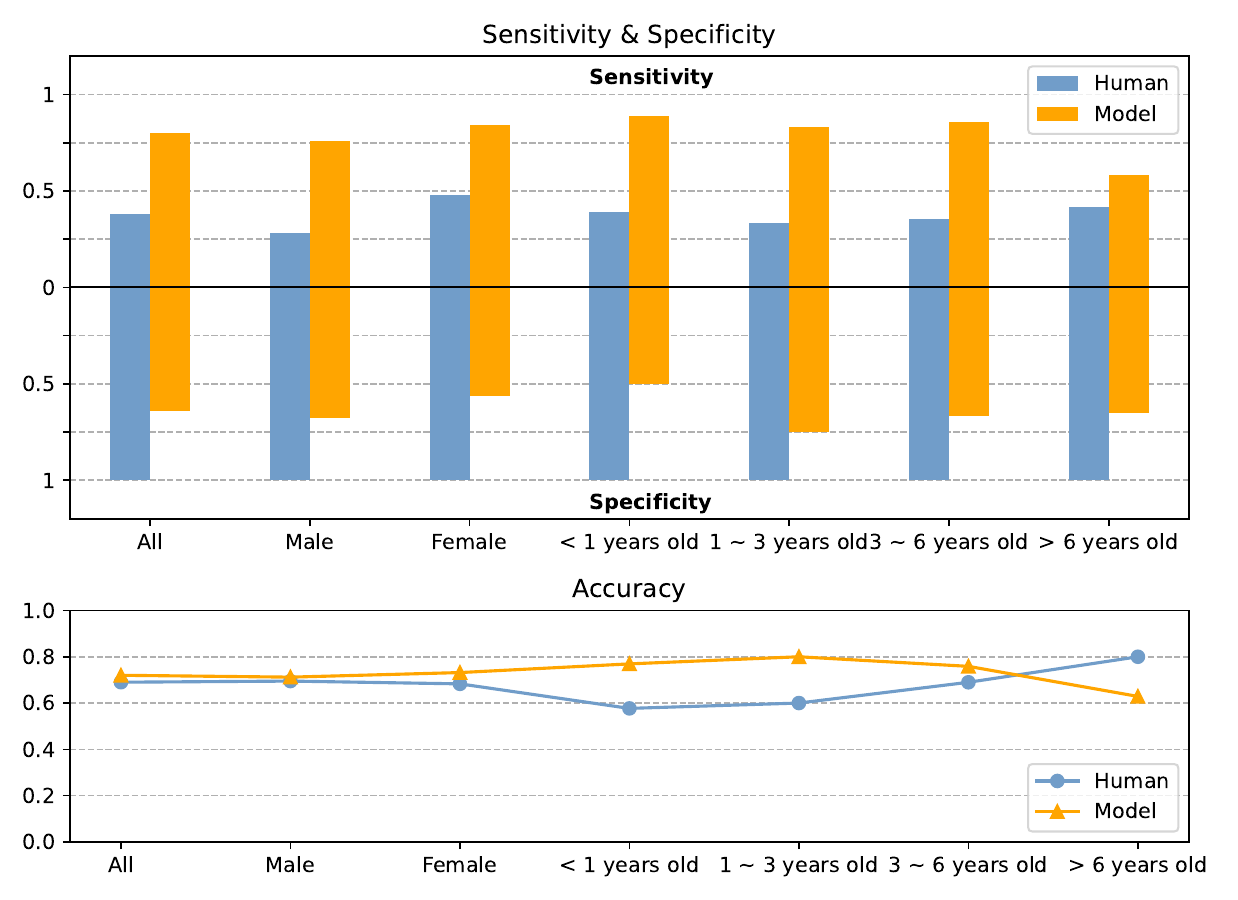}
        \label{fig:human_eval_sensitivity_specificity_accuracy}
    }
    
    \caption{The statistics of test samples and diagnostic performance comparison between human and AuscultaBase. (a) The overview of the clinical evaluation process. (b) The distribution of test samples across gender and age. (c) The confusion matrices of the diagnostic results derived by the human and AuscultaBase. (d) The sensitivity, specificity, and diagnostic accuracy across different genders and age groups. The sensitivity, specificity, and accuracy range from 0 to 1, and \textcolor{red}{higher} is better.}
    \label{fig:enter-label}
\end{figure}

We further assessed the diagnostic performance of AuscultaBase by comparing its predictions with those of experienced human clinicians in detecting congenital heart disease (CHD). For this evaluation, we collected 100 heart sound recordings.
The auscultation protocol consists of recordings over 5 body sites: aortic region (right 2nd intercostal space), pulmonic region (left 2nd intercostal space, parasternal), Erb's point (left 3rd intercostal space aka left lower sternal border), tricuspid region (left 4th intercostal space, parasternal), mitral region (left 5th intercostal space, midclavicular). The auscultation positions are visualized in Figure~\ref{fig:human_eval_process}. To detect sufficient cardiac cycles, at least 15s of heart sound using direct skin contact is obtained per site. The electronically amplified stethoscope (Littmann 3200, 3M) is used for data acquisition. During the examination the participant is seated, laid down, or held to the most comfortable position.
Additionally, each recording is paired with a cardiac ultrasound report serving as the diagnostic gold standard.
The dataset comprises recordings from 50 healthy children and 50 children diagnosed with CHD (including atrial and ventricular septal defects), with participant ages ranging from 11 days to 12 years (Figure~\ref{fig:human_eval_statistics}). Five clinicians from Anonymous Hospital, each with over 15 years of experience, independently diagnosed these recordings without access to the corresponding ultrasound reports. In parallel, the AuscultaBase model was fine-tuned on the HSDReport dataset and evaluated on these same 100 samples, ensuring that the test recordings were excluded from the fine-tuning process.

\subsubsection{Diagnostic Performance: Insights and Implications}

The results, depicted in Figures~\ref{fig:human_eval_confusion_matrix} and~\ref{fig:human_eval_sensitivity_specificity_accuracy}, yield several key insights:

\textbf{(1) Enhanced Sensitivity.}
AuscultaBase achieved a sensitivity of 80\%, substantially higher than the 38\% attained by human experts, though at the expense of a lower specificity (64\% versus 100\%). The high sensitivity of the model indicates a reduced risk of missing diseased cases, thereby enabling earlier interventions—an essential factor in CHD detection. In practice, a synergistic diagnostic approach that combines the high sensitivity of AuscultaBase with the high specificity of experienced clinicians may offer the best of both worlds, optimizing early detection while minimizing false positives.

\textbf{(2) Superior Diagnostic Accuracy in Young Patients.}
Overall, AuscultaBase achieved an accuracy of 72\%, slightly surpassing the 69\% accuracy of human experts. Notably, the model demonstrated marked superiority in younger age groups—achieving 77\% accuracy in children under one year old compared to 58\% for human experts, and 80\% accuracy in children aged one to three years versus 60\% for human experts. Early and accurate diagnosis in these critical age groups is vital; for example, timely intervention can prevent the progression of CHD-related complications such as pulmonary arterial hypertension (PAH) and subsequent Eisenmenger syndrome.

\textbf{(3) Promising Role as a Diagnostic Assistant.}
The complementary strengths of AuscultaBase and human experts suggest a promising pathway toward integrated diagnostic systems. By leveraging the model’s high sensitivity to flag potential cases and utilizing the expertise of clinicians for confirmation, the overall diagnostic process can be both more accurate and more efficient. This integrative approach could transform clinical workflows in pediatric cardiology, ultimately leading to better patient outcomes.

In summary, our experimental results demonstrate that AuscultaBase not only outperforms existing baseline models across a variety of tasks but also shows significant promise as a clinical diagnostic tool. Its robust performance in the face of data imbalance, adaptability across diverse sound types and task formulations, and its potential to complement human expertise highlight its substantial potential to improve early detection and treatment in healthcare settings.
\section{Discussion}

This work presents a large-scale body sound corpus (AuscultaCorpus), an auscultation acoustic foundation model (AuscultaBase), and a comprehensive benchmark for evaluation (AuscultaBench). Our work addresses existing limitations in traditional auscultation practices through the integration of AI. By training on a diverse collection of body sounds using contrastive learning, our model shows a substantial improvement over existing pre-trained acoustic models, especially in imbalanced and limited-data settings. The full fine-tuning results reveal the robustness of our model in both binary and multi-class classification tasks, underscoring its potential for real-world clinical applications. The comparison of the auscultation efficacy between human experts and our model also indicates our model's significance and clinical potential as a diagnostic assistant. This work emphasizes the value of combining traditional auscultation with AI-driven methods to deliver more reliable and accessible healthcare solutions, paving the way for the broader adoption of AI in medical diagnostics.

Given the cirtical need for rapid and cost-effective screening solutions, in this study, we proposed a comprehensive framework, including the AuscultaCorpus (a large-scale body sound corpus), the AuscultaBase (an auscultation acoustic foundation model), and the AuscultaBench (a comprehensive evaluation benchmark), to advance AI-assisted auscultation and improve the diagnostic capabilities of primary care providers while enabling scalable, high-quality body sound analysis. These contributions not only address the limitations of traditional auscultation practice, but also provide strong support for the application of AI in medical diagnosis.

The AuscultaCorpus offers a diverse set of body sound data, enabling the AuscultaBase model to learn more robust feature representations. This, in turn, improves performance in diverse diagnostic downstream tasks, outperforming baseline models. Notably, the model’s success validates AuscultaBase as a valuable tool for clinical practice. The AuscultaBench provides a structured framework for evaluating model performance across 16 tasks within two key diagnostic categories: abnormality detection and disease diagnosis. This ensures consistent, reproducible comparisons and drives progress in automated diagnostic methods.

AuscultaBase demonstrated strong generalization, achieving the highest Macro-F1 performance in 12 out of 16 tasks and leading in Micro-F1 for 10 tasks. This robust performance highlights the model's versatility across different auscultation tasks. By contrast, baseline models such as OPERA-CT, AudioMAE, CLAP, and PANN lagged behind in terms of task completion and accuracy. Additionally, AuscultaBase showed excellent balanced accuracy, particularly in handling imbalanced datasets (refering to the Supplementary Information file), which is crucial for real-world clinical situations. In binary classification, it surpassed baseline models in terms of ROC curves and AUC value, further emphasizing its potential for clinical screening and early disease intervention.

Multi-dimensional evaluations also demonstrated AuscultaBase’s adaptability across various acoustic environments, confirming its ability to perform well in diverse real-world clinical settings. AuscultaBase not only has the ability to perform nuanced clinical assessments, but also has the ability to impact patient care through more precise disease identification. This robust performance in different acoustic environments demonstrates the adaptability of AuscultaBase, making it a versatile tool for encountering multiple sound sources and conditions in real-world clinical applications. This versatility suggests that AuscultaBase is able to effectively generalize to different problem formulations, which is an important quality for its application in different diagnostic scenarios. 

In addition, in a comparative analysis with human experts, we further validated the clinical application prospects of AuscultaBase. In the comparison, AuscultaBase demonstrated high sensitivity, indicating a reduced risk of missed cases, thus enabling early intervention. Notably, the model showed a clear advantage in disease diagnosis accuracy in young patients, where early and accurate diagnosis is crucial for such critical age groups. AuscultaBase not only outperformed existing baseline models on a variety of tasks, but also showed significant application potential as a clinical diagnostic tool. Both the adaptability across different sound types and task formulations, and the potential to complement human expertise, highlight its great potential for improving early detection and treatment in healthcare settings.

However, some limitations may still remain in our study. Although AuscultaCorpus is large, it may not encompass all possible body sound types, which could lead to suboptimal performance when encountering rare or previously unknown sounds. Furthermore, while AuscultaBase has shown excellent performance across multiple tasks, its application in real-world clinical settings requires further validation. Additionally, the interpretability of the model needs improvement, so clinicians can better understand and trust its diagnostic results. This omission stems from the practical challenges associated with clinicians annotating precise murmur timestamps for each cardiac cycle, which consequently precluded our ability to conduct interpretability analysis on temporal prediction accuracy.

In view of the existing limitations, future research can be deepened from the following dimensions: First, continue to expand and diversify the corpus of body sound data to improve the generalization ability of the model. This includes the introduction of new acquisition tools such as smartphones and wearable devices to capture more comprehensive sound frequencies and background noises and enrich data diversity. Second, strengthen the validation and optimization of AuscultaBase in clinical settings. Through multi-center clinical trials, the performance and applicability of the model are evaluated in different medical institutions and regions. Close collaboration with clinicians is essential to jointly design trial plans, collect real clinical data, and provide timely feedback on model performance to iteratively optimize model parameters and improve diagnostic accuracy and reliability.

In conclusion, the proposed framework has great potential in the field of AI-assisted auscultation. The strong generalization ability, robustness, balanced accuracy and discrimination ability of AuscultaBase provide solid support for its wide application in clinical practice. Although our framework still has some limitations, in the future, we can continue to focus on corpus expansion and clinical validation to ensure the wide application and sustainable development of AI in medical diagnosis.

\section{Online Methods}

\subsection{AuscultaCorpus: Pretraining Datasets}

A total of 11 datasets containing cardiac, respiratory, and bowel sounds were collected to construct AuscultaCorpus, which consists of more than 40,000 recordings and 300 hours. Of the body sound datasets listed in Table~\ref{tab:pretraining_datasets}, most datasets are publicly available, except for HSDreport and XHheartSound. The latter is a private dataset collected by Anonymous Hospital and has not yet been released as open source. The recordings were captured using various digital stethoscope versions positioned on the chest, heart, and abdomen. This diverse collection enhances the model’s ability to analyze body sound frequency and rhythm, ultimately improving its clinical auscultation applications. Additionally, a subset of labeled samples is held out from the pretraining dataset for subsequent downstream benchmarking. Please refer to the Supplementary Information file for more details about the data construction, auscultation positions, and patients' metadata.
\begin{table}[]
    \centering
    \resizebox{\linewidth}{!}{
        \begin{tabular}{lcllll}
        \toprule
        \textbf{Dataset} & \textbf{\makecell[c]{Sound\\Type}} & \textbf{SR} & \textbf{\#Sample} & \textbf{Duration (s)} & \textbf{Total (h)} \\
        \midrule
        SPRSound~\citep{zhang2022sprsound} & L & 8kHz & 3554 & 11.2 {[}0.3$\sim$15.4{]} & 11.0 \\
        HF Lung~\citep{hsu2021benchmarking} & L& 4kHz & 13957 & 15.0 [15.0$\sim$15.0] & 58.2 \\
        ICBHI 2017~\citep{rocha2019open} & L  & 4$\sim$44.1kHz & 920 & 21.5 {[}7.9$\sim$86.2{]} & 5.5 \\
        Lung Sound~\citep{fraiwan2021dataset} & L  & 4kHz & 336 & 17.4 {[}5.0$\sim$30.0{]} & 1.6 \\
        \makecell[l]{Respiratory\\Database@TR}~\citep{altan2017multimedia} & L  & 4kHz & 504 & 21.7 {[}14.8$\sim$30.0{]} & 3.0 \\
        Korean~\citep{yaseen2018classification} & H & 8kHz & 1000 & 2.4 {[}1.2$\sim$4.0{]} & 0.7 \\
        Cinc 2016~\citep{clifford2016classification} & H  & 2kHz & 3240 & 22.5 {[}5.3$\sim$122.0{]} & 20.2 \\
        Circor 2022\citep{oliveira2021circor} & H  & 4kHz & 3163 & 22.9 {[}5.2$\sim$64.5{]} & 20.1 \\
        HSDReport~\citep{zhao2024hsdreport} & H  & 4$\sim$8kHz & 3660 & 63.1 {[}6.9$\sim$132.7{]} & 64.2 \\
        XHheartSound & H  & 4$\sim$8kHz & 8377 & 59.0 {[}1.5$\sim$132.7{]} & 137.0 \\
        Bowel Sound~\citep{ficek2021analysis} & B & 44.1kHz & 1606 & 2.0 [2.0$\sim$2.0] & 0.9 \\
        \midrule
        Total & - & - & 40317 & - & 322.4 \\
        \bottomrule
        \end{tabular}
    }
    \caption{Statistics of the AuscultaCorpus (SR: sampling rate; Duration: mean [min, max]), where the sound type is categorized into lung (L), heart (H), and bowel (B) sounds.}
    \label{tab:pretraining_datasets}
\end{table}

Given the heterogeneity in auscultation positions, sampling rates, and recording durations of those datasets, each recording is first resampled to 16kHz and converted to a mono channel. Subsequent processing involves trimming leading and trailing silences to enhance recording quality. The refined audio segments are then transformed into 64-bin LogMel spectrograms using a Hanning window with a 32ms stride. Finally, Min–Max Normalization is applied to standardize the spectrograms prior to their use in pretraining the auscultation foundation model.

\subsection{AuscultaBase: Auscultation Foundation Model}
\subsubsection{Training Algorithm}
Given the inherent similarities among body sound recordings, distinguishing subtle differences is crucial for detecting abnormalities. To this end, we employ a contrastive learning framework during pretraining. Specifically, two segments extracted from the same recording are designated as positive pairs, whereas segments drawn from different recordings serve as negative pairs. Given two segments $x$ and $x^\prime$, the features will first be extracted with an encoder $f(\cdot)$ from them, and then a projector $g(\cdot)$ will map the features into a lower dimensional space to obtain the final representation. To improve the quality and discriminative power of the learned representations, an extra linear layer $W$ is added on top of the whole model for the positive/negative anchor $x^\prime$. Then the bilinear similarity $s(\cdot)$ is calculated as
\begin{equation}
    s(x,x^\prime) = g(f(x))^T W g(f(x^\prime)).
\end{equation}
Given a batch of samples $\mathcal{X} = \{ x_1,x_2,...,x_N \}$, the bilinear similarity will be calculated for each sample pair, where the positive anchor of $x_i$ is represented as $x^+$, and the negative distractors are $\mathcal{X}^- (x_i)=\{ x^- | x^- \neq x^+, \forall x^- \in \mathcal{X} \}$. In this way, the loss function for a batch can be treated as a multi-class classification problem and formulated as
\begin{equation}
    \mathcal{L} = -\frac{1}{N} \sum\limits_{i=1}^{N} {\rm log} \frac{{\rm exp}(s(x_i,x^+))}{\sum_{x^- \in \mathcal{X}^-(x_i)\cup\{x^+\}} {\rm} (s(x_i,x^-))}.
\end{equation}
This approach compels the encoder to discern fine-grained acoustic variations, thereby yielding robust and generalizable features suitable for subsequent downstream analyses.

\subsubsection{Model Architecture}
In view of the variable lengths of body sound recordings, we adopt a transformer-based network, following the design principles outlined in \cite{chen2022hts}, to serve as our encoder. The extracted features are subsequently projected into a lower-dimensional space via a dedicated projection layer. Within this space, a bilinear similarity metric is computed between representations, and the optimization objective is formulated to enhance the similarity of positive pairs while reducing that of negative pairs. To manage the inherent variability in audio durations within each batch, we implement random cropping to obtain standardized training segments. Furthermore, to ensure a sufficiently challenging optimization process, separate dataloaders are established for each dataset so that all audio clips within a given batch originate from the same source; batches are then sampled randomly across these dataloaders.

\subsubsection{Training Details}
We pre-train our model using AuscultaCorpus, a combination of the 11 datasets. We apply validation with the validation sets of the datasets (e.g. SPRSound, HF Lung, ICBHI 2017, and Bowel Sound). For those datasets that do not have an existing validation set, we randomly split 10\% of the data to serve for validation. To adapt to the varied lengths of recordings, we adopt random cropping for each audio and feed them to the pertaining. according to the minimum duration of the recordings for each dataset, different crop lengths are set. Specifically, the cropped length for Korean and Bowel Sound datasets is set as 320 ms, while it is 640 ms for others. The encoded representations are projected into a 512-dimensional space, and the model is trained over 200 epochs using the Adam optimizer. The initial learning rate is 1e-04 and decays by a factor of 0.99 at each epoch. Additionally, as a usual augmentation technique, we adjust the loudness of the recording~\citep{schluter2015exploring} with a random scale factor ranging from 0.9 to 1.1. Moreover, SpecAugment~\citep{park2019specaugment} is also adopted.



\subsection{AuscultaBench: Auscultation Benchmark}

\begin{table}[]
\centering
    \resizebox{\linewidth}{!}{
        \begin{tabular}{lclccll}
        \toprule
        \textbf{ID} & \textbf{Function} & \textbf{Dataset} & \textbf{\makecell[c]{Sound\\Type}} & \textbf{\makecell[c]{Task\\Type}} & \textbf{Class Name} & \textbf{Data Distribution} \\
        \midrule
       
         T1 & AD & SPRSound & L & MC & Normal / Adventitious / Poor Quality & 2324 / 1000 / 230 \\
         T2 & AD & SPRSound & L & MC & \makecell[l]{Normal / CAS / DAS / CAS \& DAS\\/ Poor Quality} & 2324 / 368 / 480 / 152 / 230 \\
         T3 & AD & HF Lung & L & BC &Normal / Abnormal & 52444 / 29489 \\
         T4 & AD & HF Lung & L & MC & Inhalation / Exhalation / CAS / DAS & 34095 / 18349 / 13883 / 15606 \\
         T5 & AD & HF Lung & L & MC & \makecell[l]{Inhalation / Exhalation / Wheeze\\/ Stridor / Rhonchi / Crackle} & \makecell[l]{34095 / 18349 / 8457 / 686\\/ 4740 / 15606} \\
        T6 & AD & ICBHI 2017 & L & MC & \makecell[l]{Normal / Crackle / Wheeze\\/ Crackle\&Wheeze} & 3642 / 1864 / 886 / 506 \\
        T7 & AD & Lung Sound & L & MC & \makecell[l]{Normal / Crepitation / Wheeze\\/ Crackle / Bronchi / Wheeze \& Crackle\\/ Bronchi\&Crackle} & 105 / 69 / 123 / 24 / 3 / 6 / 6 \\
        T8 & AD & Circor 2022 & H & MC & Murmur Present / Absent / Unknown & 363 / 2391 / 156 \\
        T9 & AD & Bowel Sound & B & BC & Bowel sound Present / Absent & 1283 / 323 \\
        T10 & DD & ICBHI 2017 & L & MC & \makecell[l]{Healthy / Bronchiectasis\\/ Bronchiolitis / COPD\\/ Pneumonia / URTI} & 35 / 16 / 13 / 793 / 37 / 23  \\
        T11 & DD & Lung Sound & L & ML & \makecell[l]{Normal / Asthma / Pneumonia / COPD\\/ BRON / Heart failure / Lung fibrosis\\/ Pleural effusion} & 105 / 99 / 15 / 33 / 9 / 63 / 18 / 6 \\
        T12 & DD & \makecell[l]{Respiratory\\Database@TR} & L & MC & COPD0 / COPD1 / COPD2 / COPD3 / COPD4 & 72 / 60 / 84 / 84 / 204 \\
        T13 & DD & Korean & H & MC & \makecell[l]{Normal / Aortic Stenosis \\/ Mitral Regurgitation / Mitral Stenosis\\/ Murmur in Systole} & 200 / 200 / 200 / 200 / 200 \\
        T14 & DD & Cinc 2016 & H & BC & Normal / Abnormal & 2575 / 665 \\
        T15 & DD & Circor 2022 & H & BC & Normal / Abnormal & 1632 / 1531 \\
        T16 & DD & HSDReport &  H & BC & Normal / Abnormal & 247 / 2028 \\
        
        \bottomrule
        \end{tabular}
    }
    \caption{Downstream task characteristics of AuscultaBench grouped into abnormality detection (AD: T1-T9) and disease diagnosis tasks(DD: T10-T16). The task type is categorized into binary classification (BC), multi-class classification (MC), and multi-label classification (ML). Data Distribution denotes the number of samples corresponding to the class names.}
    \label{tab:tasks}
\end{table}

\subsubsection{Tasks}
Our benchmark comprises 16 sub-tasks encompassing binary classification (BC), multi-class classification (MC), and multi-label classification (ML), as detailed in Figure~\ref{fig:benchmark_sanky}. Please refer to Table~\ref{tab:tasks} for more details. These tasks are organized by their functional objectives into abnormality detection and disease diagnosis.

\paragraph{Abnormality Detection} tasks (T1-T9) are designed to discern whether a body sound is normal and, if abnormal, to specify the type of abnormality or normality present. For example, task T8 is aimed at identifying the presence of murmurs in heart sounds, whereas task T4 requires the model to determine whether the patient is inhaling or exhaling and, if adventitious sounds are detected, to further classify them as continuous (CAS) or discrete (DAS). While abnormalities in body sounds may signal potential health concerns, these tasks do not substitute for a definitive clinical diagnosis, which must be confirmed by a clinician.

\paragraph{Disease Diagnosis} tasks (T10-T16) focus on determining whether a patient is healthy or affected by a specific disease. For instance, task T13 distinguishes between healthy heart sounds and those indicative of cardiac disease, and task T12 involves identifying the stage of chronic obstructive pulmonary disease (COPD), ranging from 0 to 4. Notably, we have excluded the “LRTI” and “Asthma” categories from the original dataset for task T10 due to their minimal representation, which precludes their appearance in the test set. These tasks require the model to analyze the entire recording to detect potential disease signatures rather than relying solely on localized features, and they provide disease-specific diagnoses rather than preliminary symptomatic observations.

\subsubsection{Data Pre-processing}
we adopt the validation set of the pretraining stage as the test set, which is always held out for pretraining and downstream fine-tuning. Considering the varied lengths of different recordings even within the same dataset, each audio in the training set is chunked and padded into the same length. Specifically, each recording of all the datasets is chunked into 8s to satisfy the input requirements of different models except for T3-T5 (HF Lung dataset), T9 (Bowel Sound dataset), T13 (Korean dataset), and T16 (HSDReport dataset). As for the HF Lung, Bowel Sound, Korean, and HSDReport datasets, the maximum durations are set as 2s, 4s, 75s, and 2s respectively because their recordings are rather short or long as demonstrated in Table~\ref{tab:pretraining_datasets}.

\begin{table}[]
    \centering
    \begin{tabular}{lccccc}
        \toprule
        &\textbf{ OPERA-CT} & \textbf{AudioMAE} & \textbf{CLAP} & \textbf{PANN} & \textbf{AuscultaBase} \\
        \midrule
        \# Parameters (M) & 31 & 86 & 80 & 82 & 31  \\
        Feature Dim. & 768 & 768 & 1024 & 2048 & 768 \\
        \# Samples (K) & 136 & 2000 & 128 & 1934 & 40 \\
        Total Duration (h) & 440 & 5000+ & $\sim$250 & 5000+ &  322 \\
        \bottomrule
    \end{tabular}
    \vspace{5pt}
    \caption{Comparison of existing acoustic pre-trained models and AuscultaBase.}
    \label{tab:comparison_models}
\end{table}

\subsubsection{Baselines}
\paragraph{OPERA-CT} \citep{zhang2024towards} is a contrastive learning-based pre-trained model designed specifically for respiratory audio feature extraction. As one of the first respiratory acoustic foundation models, the OPERA family leverages extensive collections of breathing and cough sounds to pretrain its architecture, thereby demonstrating superior performance and generalizability compared to conventional acoustic models pre-trained on general audio. In this study, we adopt OPERA-CT, the variant with the largest number of parameters, as a baseline.

\paragraph{AudioMAE}\citep{huang2022masked} is a self-supervised pre-trained model for audio understanding, inspired by the Masked Autoencoders (MAE) approach originally developed for vision tasks \citep{he2022masked}. It operates by randomly masking segments of the audio signal during pretraining and then learning to reconstruct the missing portions. This strategy enables the model to capture salient audio features without reliance on labeled data, and it has demonstrated robust performance across a range of audio classification and sound event detection tasks.

\paragraph{CLAP}\citep{elizalde2023clap} is a language-supervised pre-trained model that bridges audio and natural language via contrastive learning. By training on paired audio and text data, CLAP aligns audio features with their corresponding textual descriptions, facilitating cross-modal retrieval and classification. This approach enhances the model’s ability to interpret and relate audio content to high-level semantic information, making it a strong baseline for audio-linguistic tasks.

\paragraph{PANN}\citep{kong2020panns} is a convolutional neural network tailored for audio pattern recognition and pre-trained on large-scale datasets for audio tagging and sound event detection. It achieves high accuracy in detecting environmental sounds, musical instruments, and other acoustic events due to its efficient architecture for audio feature extraction. Further comparisons concerning the number of parameters, feature dimensions, total sample counts, and the aggregate durations of recordings in the pretraining datasets are detailed in Table~\ref{tab:comparison_models}.

Within our benchmark of downstream tasks, we have chosen four pre-trained acoustic models to compare with our models: OPERA-CT, CLAP, AudioMAE, and PANN. The official implementations for each baseline are strictly followed. For example, we resampled all of the recordings with 44.1 kHz and 32 kHz for CLAP and PANN respectively.

\subsubsection{Benchmarking Protocol}
All tasks are evaluated using the standard linear probe protocol, which freezes the pre-trained audio encoder and only trains a single fully connected layer on top of the encoded representations. Such a protocol focuses on the quality and generalizability of the encoded features and avoids the overfitting problem with a small dataset. It is noted that we adopt different data pre-processing strategies (e.g. resampling and feature extraction) for different baselines strictly following their own official implementations. As for our foundation model, the same audio pre-processing procedures are applied as in the pertaining stage. Due to the significant variation in audio lengths (e.g. 5.3$\sim$122.0 seconds for Cinc~2016), the training set recordings are segmented into equal-length portions prior to training. The linear head is then fine-tuned for 64 epochs using a decaying learning rate, with an initial value of 1e-4.

\subsubsection{Evaluation Metric}

For performance evaluation, Macro-F1 and Micro-F1 are reported respectively for each classification task to evaluate the performance across classes and samples. We report both of them because the Macro-F1 scores reflect the model's ability to handle classes equally well, irrespective of class size, while the Micro-F1 scores emphasize overall classification accuracy across all classes. Additionally, Precision and Recall scores are also reported for each task. The ROC (Receiver Operating Characteristic) curve and AUC (Area Under the ROC curve) are also plotted and calculated for binary classification tasks.
For a comprehensive overall evaluation, we also report the BCS (Borda Count Score) of performance categorized by different perspectives.

For comparative analysis between our model and human experts, we compared the diagnostic outcomes using standard metrics: confusion matrices, sensitivity, specificity, and overall accuracy. The confusion matrix provides a detailed breakdown of true positives (TP), false positives (FP), false negatives (FN), and true negatives (TN). Sensitivity reflects the ability to correctly identify diseased cases, while specificity indicates the correct identification of healthy cases. Accuracy, the overall proportion of correct predictions, provides a holistic measure of diagnostic performance. These metrics are formulated as:
\begin{equation}
    \small
    Sensitivity=\frac{TP}{TP+FN}, Specificity=\frac{TN}{FP+TN} , Accuracy=\frac{TP+TN}{TP+FP+FN+TN}.
    \label{eq:sensitivity_specificity_accuracy}
\end{equation}
\section*{Ethical Considerations}
This research was conducted in accordance with established ethical principles for medical studies, ensuring the rights, privacy, and well-being of all participants. Below is an overview of the primary ethical considerations addressed.

\paragraph{Informed Consent}
Prior to any data collection, all participants or their legal representatives received detailed information about the objectives, methods, and potential applications of this study. Written informed consent was obtained from every participant, aligning with international standards such as the Declaration of Helsinki. This process guaranteed that individuals were fully aware of both the scope and implications of their involvement in the research.

\paragraph{Data Confidentiality and Security}
To protect participant privacy, strict measures were implemented to remove any personal identifiers from collected data, ensuring anonymity. Additionally, encrypted storage systems with controlled access were employed to maintain data security and prevent unauthorized viewing or manipulation. These steps were critical in upholding the confidentiality of the information throughout the study.

\paragraph{Ethical Review and Oversight}
This study protocol underwent a comprehensive evaluation by an independent ethics committee, receiving approval under the reference number XHEC-C-2020-009-1. The committee provided ongoing oversight to confirm adherence to ethical standards and safeguard participant welfare at every stage of the project.

\section*{Data Availability}
We utilized 11 datasets in our benchmark, with access methods and licensing details provided in Extended Table~\ref{tab:dataset_availability}. Notably, most of datasets include both audio recordings and accompanying metadata. The audio data has been anonymized, and the metadata contains no personally identifiable information or offensive content.

\section*{Code Availability}
The code and model checkpoint has been released in \url{https://github.com/applewpj/AuscultaBase}.

\bibliography{reference}
\bibliographystyle{iclr2025_conference}

\appendix

\newpage

\etocdepthtag.toc{mtappendix}
\etocsettagdepth{mtchapter}{none}
\etocsettagdepth{mtappendix}{subsection}
\tableofcontents

\section{Related Works}
The auscultatory body sounds encompass heart, respiratory, and bowel sounds, with most existing models focusing on a single category. For heart sounds, the primary tasks include abnormality detection and disease diagnosis. Traditional methods often rely on acoustic features such as STFT for feature processing. For instance, \cite{chen2023robust} applied noise reduction followed by STFT to obtain spectrograms, using a CNN with an attention module for classification. Similarly, \cite{guo2023ds} enhanced feature extraction by combining high-order spectral estimation with STFT, employing a dual-stream CNN as the primary architecture. General audio pre-training models have also been explored. For example, \citep{koike2020audio} compared PANN-extracted features \citep{kong2020panns} with STFT-derived features used in non-pre-trained architectures like VGG and ResNet, demonstrating the transferable capabilities of pre-trained models in heart sound diagnosis.

Respiratory sound analysis similarly focuses on abnormality detection and disease diagnosis. Early approaches utilized traditional feature extraction techniques such as STFT and MFCC \citep{oletic2017asthmatic, jung2021efficiently}. Recently, the COVID-19 pandemic spurred the development of pre-training models for respiratory sounds. \cite{zhang2024towards} constructed a dataset combining cough and stethoscope-collected respiratory sounds, comparing contrastive and generative pre-training methods across 19 tasks. Additionally, \cite{baur2024hear} developed an acoustic event detector to identify respiratory-related events from online audio sources, leveraging a generative pre-training approach based on MAE \citep{he2022masked}.

Bowel sounds remain underexplored due to a lack of datasets and established methods. Current work focuses primarily on bowel sound recognition. For instance, \cite{sitaula2022neonatal} collected data from 49 newborns and developed a hybrid CNN and hidden semi-Markov model using MFCC features. Similarly, \cite{baronetto2023segment} investigated the utility of AudioSet \citep{gemmeke2017audio} for pre-training and evaluated CNN and attention-based architectures on data from 18 participants.

While these efforts represent progress, they fail to address a critical gap: the intrinsic inter-relatedness of body sounds in auscultation. Existing methods operate in silos, treating each body sound category independently and overlooking shared patterns and synergies. Furthermore, no existing approach attempts to create a unified framework capable of leveraging the collective knowledge of all auscultation data.

To overcome these limitations, we introduce AuscultaBase, the first-ever audio foundation model designed specifically for body sound auscultation. Inspired by the transformative success of foundation models in vision and language domains, AuscultaBase introduces a multi-layered knowledge integration paradigm that unifies data across heart, respiratory, and bowel sound categories. It achieves this through an adaptive acoustic fusion mechanism, enabling dynamic representation learning from diverse audio sources.

\section{Datasets}


\subsection{Datasets Availability}

\begin{table}[htbp]
    \centering
    \resizebox{\linewidth}{!}{
    \begin{tabular}{llll}
    \toprule
        \multicolumn{1}{c} {\textbf{Dataset}} & \multicolumn{1}{c} {\textbf{Source}} & \multicolumn{1}{c} {\textbf{Access}} & \multicolumn{1}{c}{\textbf{License}} \\
         \midrule
            SPRSound & SJTU & 	
\href{https://github.com/SJTU-YONGFU-RESEARCH-GRP/SPRSound}{https://github.com/SJTU-YONGFU-RESEARCH-GRP/SPRSound} & CC-BY-4.0 \\
         \midrule
            HF Lung & NTU & \href{https://gitlab.com/techsupportHF/HF_Lung_V1}{https://gitlab.com/techsupportHF/HF\_Lung\_V1} & CC-BY-4.0 \\
         \midrule
            ICBHI 2017 & * & \href{https://bhichallenge.med.auth.gr/ICBHI_2017_Challenge}{https://bhichallenge.med.auth.gr/ICBHI\_2017\_Challenge} & CC0 \\
                 \midrule
            Lung Sound & JUST\&KAUH & \href{https://data.mendeley.com/datasets/jwyy9np4gv/3}{https://data.mendeley.com/datasets/jwyy9np4gv/3} & CC-BY-4.0 \\
         \midrule
        RespiratoryDatabase@TR & ITU & \href{https://data.mendeley.com/datasets/p9z4h98s6j/1}{https://data.mendeley.com/datasets/p9z4h98s6j/1} & CC-BY-4.0 \\
         \midrule
            Korean & SJU & \href{https://github.com/yaseen21khan/Classification-of-Heart-Sound-Signal-Using-Multiple-Features-}{https://github.com/yaseen21khan/Classification-of-Heart-Sound-Signal-Using-Multiple-Features-} & CC-BY-4.0 \\
                 \midrule
            Cinc 2016 & * & \href{https://archive.physionet.org/physiobank/database/challenge/2016/}{https://archive.physionet.org/physiobank/database/challenge/2016/} & Custom license \\
         \midrule
            Circor 2022 & * & \href{https://physionet.org/content/circor-heart-sound/1.0.3/}{https://physionet.org/content/circor-heart-sound/1.0.3/} & ODC-By \\
         \midrule
            HSDReport & SJTU & not available & - \\
                     \midrule
            XHheartSound & - & - & - \\
                                 \midrule
            Bowel Sound & WUT\&PUMS & \href{https://www.kaggle.com/robertnowak/bowel-sounds}{https://www.kaggle.com/robertnowak/bowel-sounds} & CC BY-NC 4.0 \\
        \bottomrule
    \end{tabular}
    }
    \caption{Dataset availability. *The Cinc 2016, Circor 2022, and HF Lung datasets are derived from multiple sources, as detailed in the text descriptions below. The datasets presented in the HSDReport are not yet publicly available.}
    \label{tab:dataset_availability}
\end{table}

\subsection{Datasets Description}

\paragraph{SPRSound~\citep{zhang2022sprsound}.}
The database is the first open-access pediatric respiratory sound database, jointly developed by Shanghai Jiao Tong University and its affiliated hospitals, with the aim of analyzing respiratory sounds in children. It comprises 2,683 recordings and 9,089 respiratory sound events from 292 participants, with a total duration of 11 hours. It includes annotations at both the event and record levels.
At the event level, the distribution of sound types is as follows: Normal (6,887), Rhonchi (53), Wheeze (865), Stridor (17), Coarse Crackle (66), Fine Crackle (1,167), and Wheeze \& Crackle (34). At the record level, the number of recordings categorized into Normal, Continuous Adventitious Sounds (CAS), Discrete Adventitious Sounds (DAS), CAS \& DAS, and Poor Quality are 1,785, 233, 347, 131, and 187, respectively. The average duration of respiratory sound events and records is 1.3 seconds and 11 seconds, respectively.
The sounds were recorded using a digital stethoscope (Yunting Model II) at four back locations: left posterior, left lateral, right posterior, and right lateral. Recordings at each location lasted over 9 seconds to capture at least two respiratory cycles (one cycle consisting of inhalation and exhalation). 

\paragraph{HF Lung~\citep{hsu2021benchmarking}.}
This lung sound database was designed to facilitate the development of algorithms for detecting inhalation, exhalation, and adventitious lung sounds. The recordings originate from two primary sources. The first source is a database used during the 2020 Taiwan Smart Emergency and Critical Care (TSECC) datathon, licensed under the Creative Commons Attribution 4.0 (CC BY 4.0) license, and provided by the Taiwan Society of Emergency and Critical Care Medicine (TSECCM). This dataset includes lung sound recordings from 261 patients.
The second source consists of recordings from 18 residents of a respiratory care ward (RCW) or respiratory care center (RCC) in Northern Taiwan, collected between August 2018 and October 2019. These recordings were approved by the Research Ethics Review Committee of Far Eastern Memorial Hospital (Case No. 107052-F), and written informed consent was obtained from all participants.
The database contains 9,765 lung sound recordings, each 15 seconds in length, and includes corresponding labels: 34,095 inhalation labels, 18,349 exhalation labels, 13,883 continuous adventitious sound labels (8,457 wheeze, 686 stridor, 4,740 rhonchi), and 15,606 discontinuous adventitious sound labels (all classified as crackles). The database, HF\_Lung\_V1, developed by Heroic-Faith Medical Science Co. Ltd., is licensed under a Creative Commons Attribution 4.0 (CC BY 4.0) International License. All patients were Taiwanese and over 20 years of age.

\paragraph{ICBHI 2017~\citep{rocha2019open}.}
The International Conference on Biomedical and Health Informatics (ICBHI) 2017 database comprises 920 respiratory sound recordings from 126 participants, along with two sets of annotations. One set includes 6,898 annotated respiratory cycles, identifying the presence of crackles, wheezes, both, or no adventitious sounds. The second set provides detailed annotations for 10,775 crackle and wheeze events.
The database contains audio samples independently collected by two research teams over several years in different countries, with ethical approval obtained from the relevant institutions. Most recordings were contributed by the Respiratory Research and Rehabilitation Laboratory (Lab3R) at the University of Aveiro, Portugal. These samples were gathered from various clinical settings, including Hospital Infante D. Pedro in Aveiro, Hospital Santa Maria and Lusíadas in Porto, and the University of Southampton, UK. Five separate studies from this team are represented, with recordings from the trachea and six chest locations (left and right anterior, posterior, and lateral) in both clinical and home environments. Participants ranged in age and included individuals with conditions such as lower and upper respiratory tract infections, pneumonia, COPD, asthma, bronchiolitis, bronchiectasis, and cystic fibrosis.

\paragraph{Lung Sound~\citep{fraiwan2021dataset}.}
The dataset contains respiratory sound recordings from 112 subjects, including 35 healthy individuals and 77 with various respiratory conditions. The subjects' ages ranged in from 21 to 90 years, with a mean age of 50.5 ± 19.4 years, and the group comprises 43 females and 69 males. Each recording lasts between 5 and 30 seconds, which is sufficient to capture at least one full respiratory cycle.
Lung sounds were recorded using an electronic stethoscope, with the stethoscope placement determined by a specialist physician. The dataset includes normal breathing sounds as well as sounds associated with seven conditions: asthma, heart failure, pneumonia, bronchitis, pleural effusion, lung fibrosis, and chronic obstructive pulmonary disease (COPD). Each recording was made three times, using different frequency filters to highlight various bodily sounds.

\paragraph{RespiratoryDatabase@TR~\citep{altan2017multimedia}.}
The RespiratoryDatabase@TR is a comprehensive database designed to facilitate the analysis of respiratory disorders. Developed by Iskenderun Technical University, Turkey. It includes lung and heart sounds, chest X-rays, pulmonary function test (PFT) data, and responses to the St. George Respiratory Questionnaire (SGRQ-C). These data were collected from patients at Antakya State Hospital using digital stethoscopes to capture lung sounds from 12 channels across the upper, middle, and lower lung regions, as well as the costophrenic angles. Heart sounds were recorded from four key areas: aortic, pulmonary, tricuspid, and mitral regions.
The database, validated and labeled by two pulmonologists, categorizes subjects based on chest X-rays, PFT results, and auscultation findings, classifying them into five levels of COPD severity (COPD0 to COPD4). It includes both healthy individuals and patients with respiratory conditions such as asthma, COPD, and bronchitis, offering a broad spectrum of clinical insights into respiratory health.

\paragraph{Korean~\citep{yaseen2018classification}.}
The database, released by Sejong University, consists of two sets: a normal set and an abnormal set. It is further divided into five categories: one normal category (N) and four abnormal categories—aortic stenosis (AS), mitral stenosis (MS), mitral regurgitation (MR), and mitral valve prolapse (MVP). Each category contains 200 audio files, for a total of 1,000 files, all in ".wav" format.
The data were collected from various sources, including auscultation skill CDs, heart sound educational materials, and 48 different websites (e.g., those from Washington, Texas, 3M, and Michigan). After removing files with excessive noise, the recordings were resampled to a frequency rate of 8,000 Hz and converted to mono channel, producing 3-period heart sound signals.

\paragraph{Cinc 2016~\citep{clifford2016classification}.}
The PhysioNet/Computing in Cardiology (CinC) Challenge 2016 database was compiled from eight sources by seven research groups worldwide. It includes nearly 30 hours of recordings, comprising 4,430 heart sound recordings from 1,072 subjects. These recordings capture 233,512 heart sounds from both healthy individuals and patients with heart valve disease or coronary artery disease, with lengths varying from a few seconds to several minutes.
The database also provides information on the subjects (age, gender), details about each recording (location on the body, duration), and additional signals such as ECG, along with data on sampling frequency and sensor type.
Heart sounds are classified into three categories: normal (no need for further diagnosis), abnormal (needs further diagnosis), and unsure (too noisy to assess). The sounds are further labeled as normal for healthy subjects or abnormal for patients with heart conditions like valve defects or coronary artery disease, including cases like mitral valve prolapse and aortic stenosis.

\paragraph{Circor 2022~\citep{oliveira2021circor}.}
This pediatric heart sound dataset was collected during two mass screening campaigns, called “Caravana do Coração” (Caravan of the Heart), conducted in the state of Paraíba, Brazil, in 2014 (CC2014) and 2015 (CC2015). Data collection was approved by the ethics board of the Complexo Hospitalar HUOC/PROCAPE, under the request of the Real Hospital Português de Beneficência.
The dataset omprises 5,282 recordings from 1,568 participants with 50.2\% male and 49.8\% female, covering four main auscultation sites. The mean age of participants was 73.4 ± 0.1 months, ranging from 0.1 to 356.1 months. In total, 215,780 heart sounds were manually annotated, with 103,853 sounds (51,945 S1 and 51,908 S2) from CC2014 and 111,927 sounds (56,449 S1 and 55,478 S2) from CC2015.
All participants completed socio-demographic questionnaires, followed by clinical examinations, nursing assessments, and cardiac investigations, including chest X-rays, electrocardiograms, and echocardiograms.

\paragraph{HSDReport~\citep{zhao2024hsdreport}.}
The dataset consists of 2,275 participants under 18 years old of age, along with 2,275 auscultation recordings and corresponding echocardiography data. The authors later expanded the dataset to include a total of 3,660 samples.
Heart sounds were recorded at five body sites: the aortic region (right 2nd intercostal space), pulmonic region (left 2nd intercostal space), Erb’s point (left 3rd intercostal space), tricuspid region (left 4th intercostal space), and mitral region (left 5th intercostal space, midclavicular line). Each site was recorded for at least 15 seconds using an electronically amplified stethoscope (Littmann 3200, 3M) in direct contact with the skin.
This study adhered to strict ethical standards, ensuring participant confidentiality and compliance with relevant regulations. Informed consent was obtained from all participants or their legal guardians, with a clear explanation of the study's purpose, the data to be collected (heart sounds and ultrasound reports), and its intended research use.

\paragraph{XHheartSound.} This dataset consists of 8, 377 auscultation audios recorded by 8, 377 children participants without diagnostic labels. The auscultation devices, positions, and procedures are followed with those of HSDReport. This research is carried out with a strong focus on ethics, protecting participant privacy, and following all rules. Before taking part, all individuals or their legal guardians were given full details about the study's goals, what data would be collected (such as heart sounds and ultrasound results), and how this data would be used for research. Consent was obtained only after making sure everyone understood these details. This dataset is collected by Anonymous Hospital, and due to the copyright constraint, it has not been open source yet.

\paragraph{Bowel Sound~\citep{ficek2021analysis}.}
The dataset, released by Warsaw University of Technology, Warsaw University, and Poznań University of Medical Sciences, is a bowel sound dataset containing 321,000 recordings, each 10 milliseconds long and labeled by medical doctors. It includes nearly 100 recordings collected by a team from the Department of Pediatric Gastroenterology and Metabolic Diseases in Poznań, Poland.
Recordings from 19 subjects were divided into 2-second fragments and then mixed. Of these, 15\% were randomly selected as test data, while the remaining 85\% were used for training and fivefold cross-validation.

\section{Demographic and Clinical Information}

\subsection{Age and Gender}
\begin{figure}
    \centering
    \includegraphics[width=\linewidth]{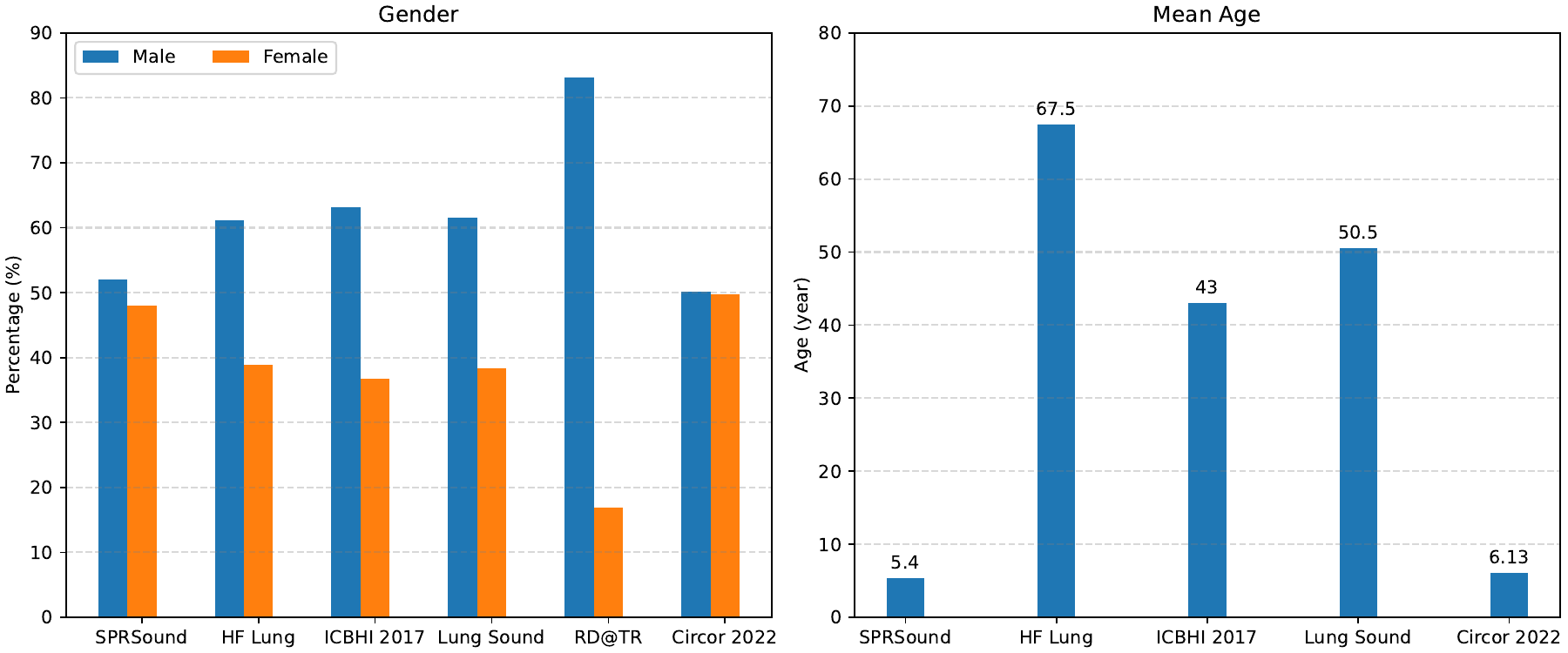}
    \caption{Gender distribution and mean age distribution for some of the datasets used in our benchmark.}
    \label{fig:age_gender_curve}
\end{figure}

\begin{figure}
    \centering
    \includegraphics[width=\linewidth]{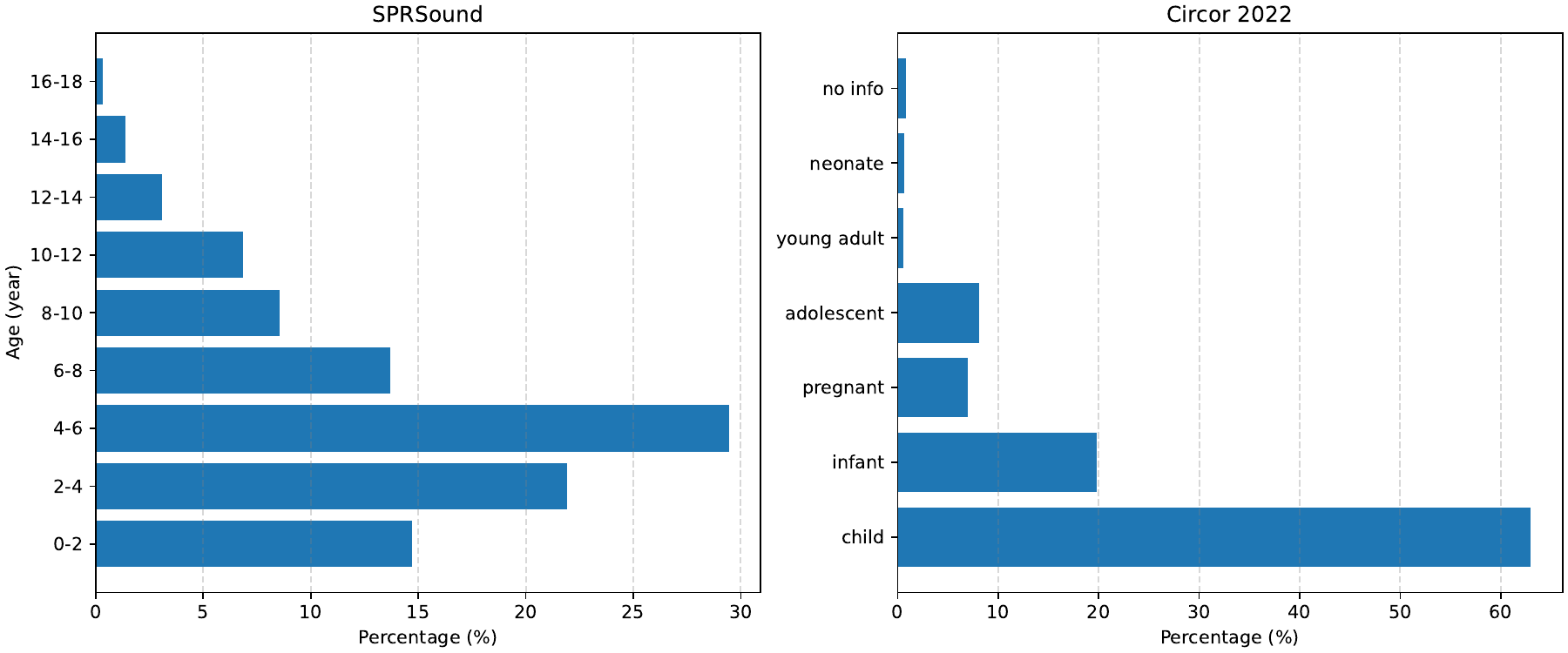}
    \caption{Segmented age distribution of the SPRSound dataset (left) and the Circor 2022 dataset (right).}
    \label{fig:age_distribution_curve}
\end{figure}

In our benchmark, we used a total of 10 datasets. The diversity and representativeness of the training data  are crucial for developing a generalizable model. We examined the population distribution of the 10 datasets used for model pre-training. The bar charts in Figure~\ref{fig:age_gender_curve} illustrate the gender distribution of 6 datasets and the mean age distribution of 5 datasets. Although the age distribution of the HSDReport dataset is unknown, it is known that all participants are children aged 18 years or younger. The distribution of the other datasets remains unknown because their details have not been published or they consist of data from multiple sources. It should be noted that the HF Lung dataset includes data from only 18 participants, as the remaining 261 data points are missing, according to the original paper. Figure~\ref{fig:age_distribution_curve} shows the segmented age distribution for the SPRSound and Circor 2022 datasets. The age categories in the Circor 2022 dataset are based on the pediatric terminology provided by the National Institute of Child Health and Human Development (NICHD).

\subsection{Auscultation position}
In our benchmark, three distinct types of data (heart sounds, lung sounds, and bowel sounds) are utilized. Below, we summarize and describe the auscultation positions used by different heart sound and lung sound datasets when collecting data. As shown in Table~\ref{tab:auscultation_position_data}, the number of auscultation positions for each heart sound dataset, along with the corresponding diagrams, is provided. The following section offers a detailed description of the auscultation positions used in the different datasets.
\begin{table}[htbp]
    \centering
    \resizebox{\linewidth}{!}{
    \begin{tabular}{lcccl}
    \toprule
        \multicolumn{1}{c} {\textbf{Dataset}} & \multicolumn{1}{c} {\textbf{Sound Type}} & \multicolumn{1}{c} {\textbf{Auscultation positions}} & \multicolumn{1}{c}{\textbf{Subfigure}} & \multicolumn{1}{c}{\textbf{Comment}}\\
         \midrule
            SPRSound & L &         
4 & (a) & \\
         \midrule
            HF Lung & L & 8 & (d) & \\
         \midrule
            ICBHI 2017 & L & 6 & (e) & It is collected from two different sources. \\
                 \midrule
            Lung Sound & L & 10 & (c) &  \\
         \midrule
        RespiratoryDatabase@TR & L+H & 16 & (f) & 4 channels for heart sounds and 12 channels for lung sounds. \\
         \midrule
            Korean & H & - & - & It is collected from different sources, and there is no fixed auscultation position. \\
                 \midrule
            Cinc 2016 & H & - & - & It is a collection of 9 heart sound datasets, without a unified standard. \\
         \midrule
            Circor 2022 & H & 4 & (b) &  \\
         \midrule
            HSDReport & H & 5 & - & \\
        \bottomrule
    \end{tabular}
    }
    \caption{Auscultation positions of different datasets and their corresponding diagram labels.}
    \label{tab:auscultation_position_data}
\end{table}

\begin{figure}
    \centering
    \subfigure[SPRSound]{
        \begin{minipage}[t]{0.25\linewidth}
        \includegraphics[width=\linewidth]{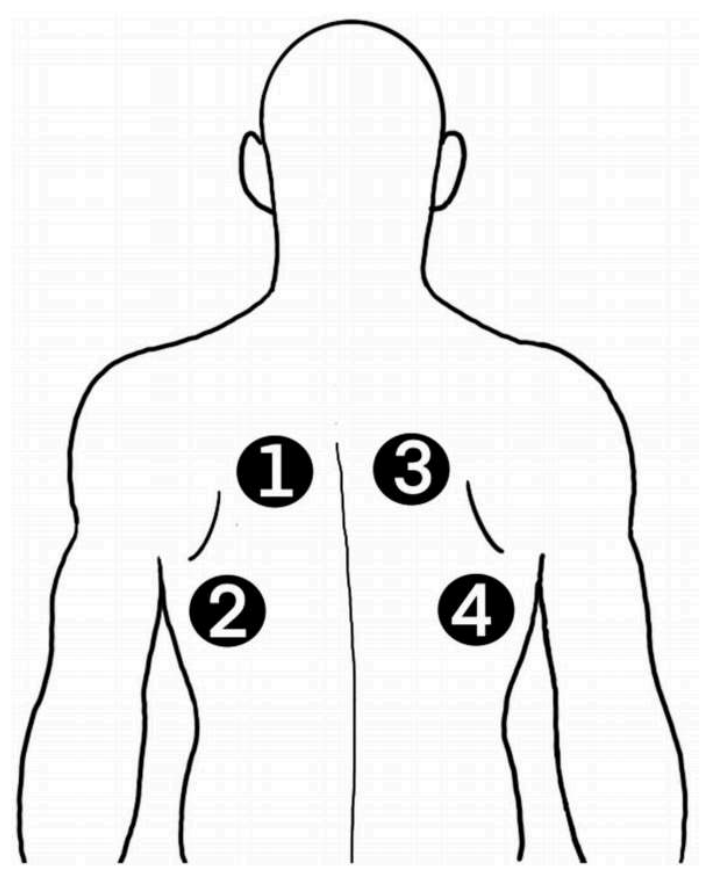}
        \end{minipage}
        \label{subfig:sprsound}
    }
    \subfigure[Circor 2022]{
        \begin{minipage}[t]{0.3\linewidth}
        \includegraphics[width=\linewidth]{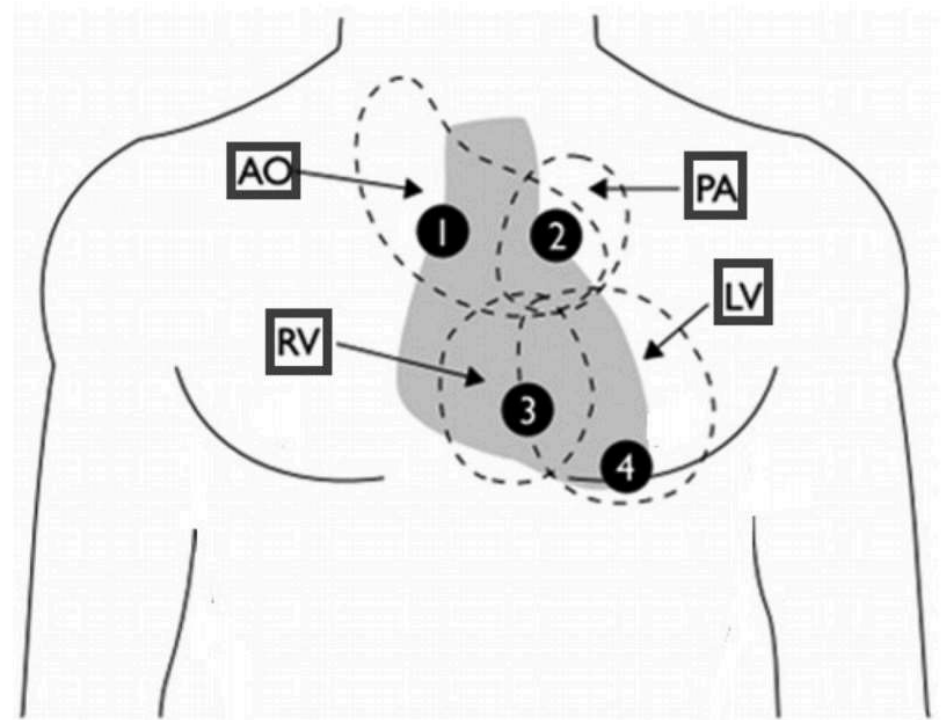}
        \end{minipage}
        \label{subfig:circor2022}
    }
    \subfigure[Lung Sound]{
        \begin{minipage}[t]{0.25\linewidth}
        \includegraphics[width=\linewidth]{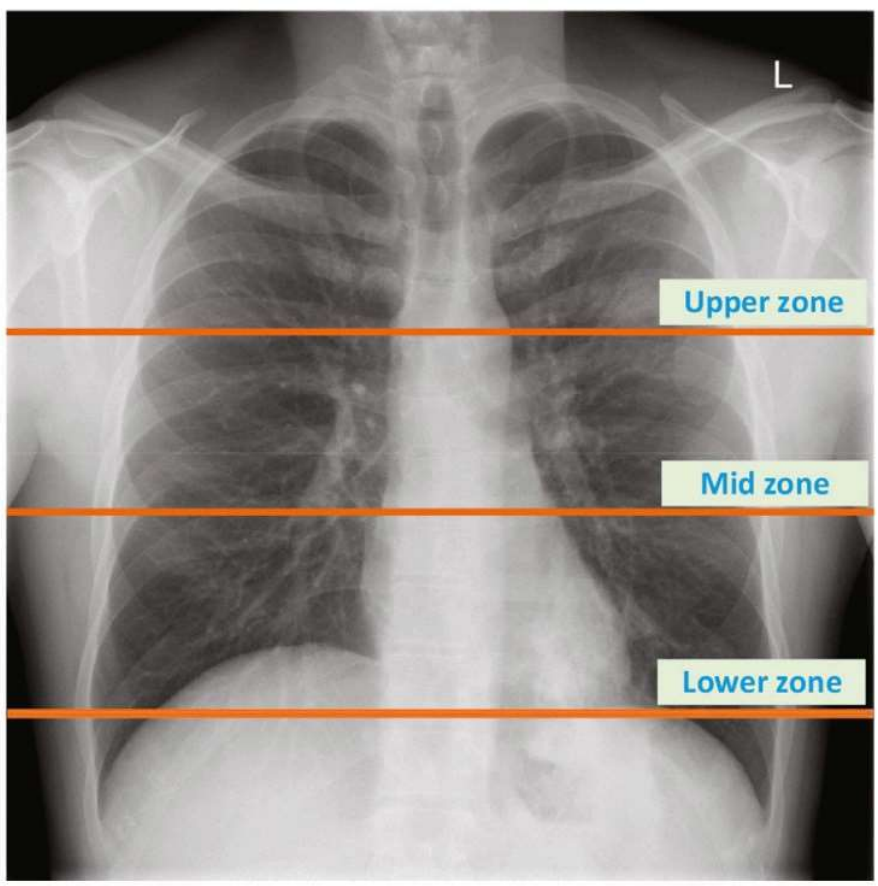}
        \end{minipage}
        \label{subfig:lung_sound}
    }
    \subfigure[HF Lung]{
        \begin{minipage}[]{0.7\linewidth}
        \includegraphics[width=\linewidth]{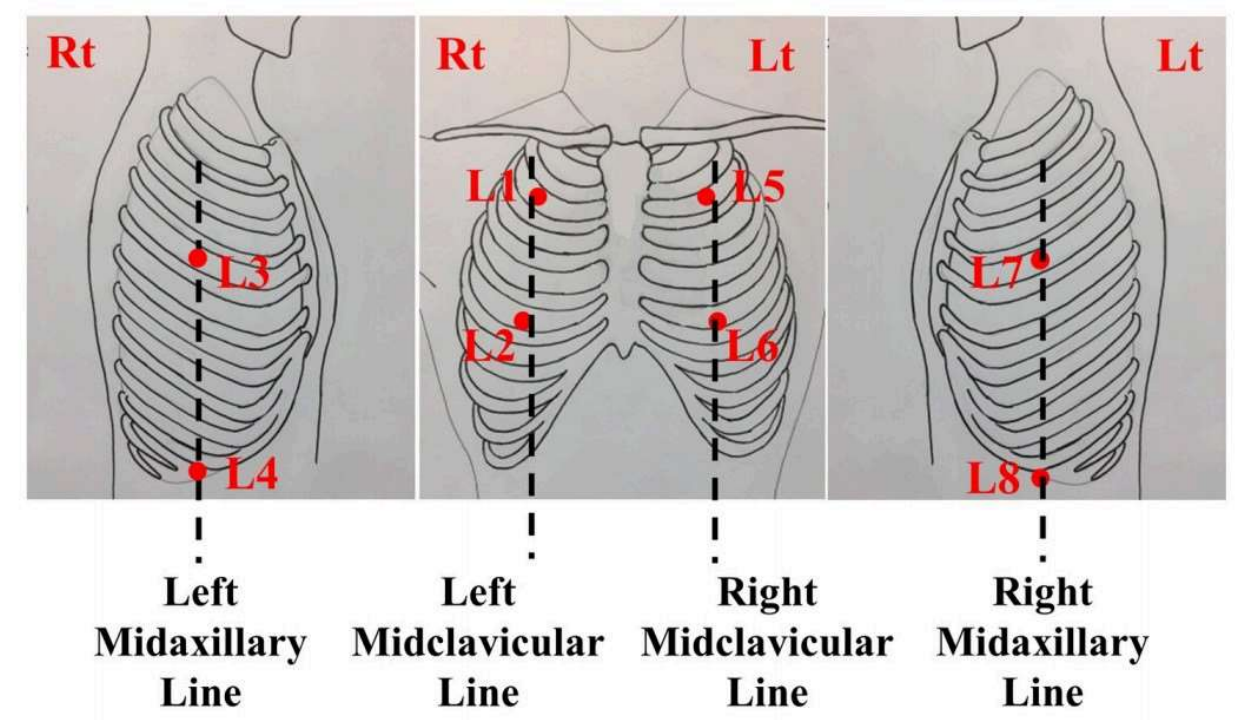}
        \end{minipage}
        \label{subfig:hf_lung}
    }
    \subfigure[ICBHI 2017]{
        \begin{minipage}[]{0.8\linewidth}
        \includegraphics[width=\linewidth]{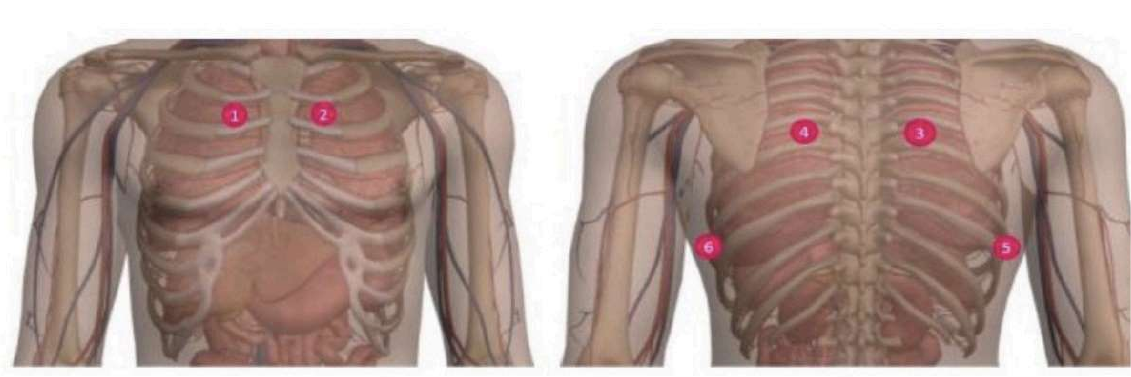}
        \end{minipage}
        \label{subfig:icbhi2017}
    }

    \subfigure[RespiratoryDatabase@TR]{
        \begin{minipage}[]{0.8\linewidth}
        \includegraphics[width=\linewidth]{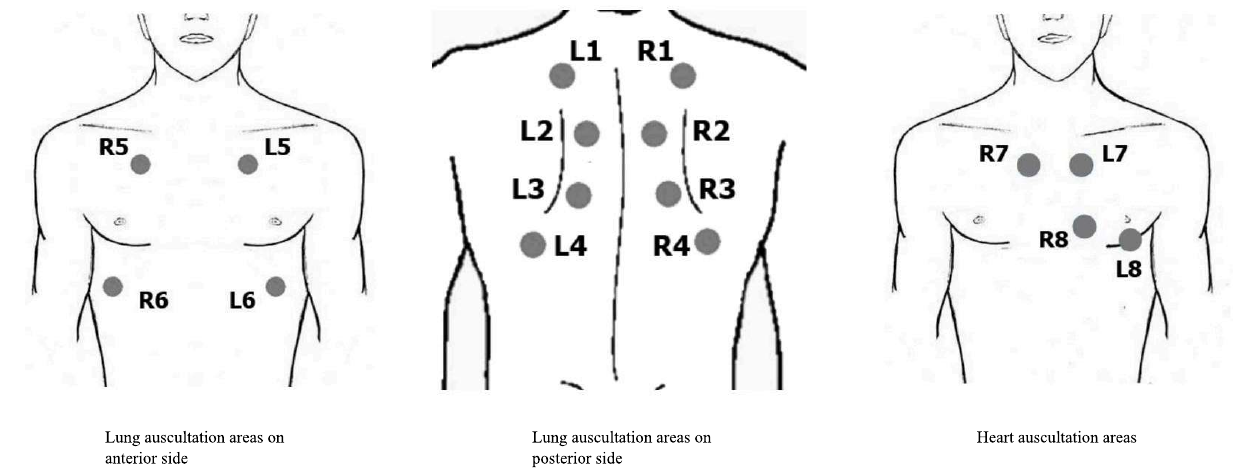}
        \end{minipage}
        \label{subfig:rd@tr}
    }

    \caption{Schematic diagram of auscultation positions for different datasets. These subfigures are all from the corresponding papers. The Circor 2022~\citep{oliveira2021circor} dataset is a heart sound dataset, RD@TR~\citep{altan2017multimedia} contains heart sound and lung sound data, and the remaining datasets (SPRSound~\citep{zhang2022sprsound},Lung Sound~\citep{fraiwan2021dataset},HF Lung~\citep{hsu2021benchmarking} and ICBHI 2017~\cite{rocha2019open}) are all composed of lung sound data.}
    \label{fig:auscultation_position}
\end{figure}

\paragraph{SPRSound.}
The respiratory sounds were recorded at four locations on the back, including left posterior, left lateral, right posterior, and right lateral as shown in Figure~\ref{subfig:sprsound}.

\paragraph{HF Lung.}
As shown in Figure~\ref{subfig:hf_lung}, there are eight auscultation positions (L1-L8) are defined as follows: L1: the second intercostal space (ICS) on the right midclavicular line (MCL); L2: the fifth ICS on the right MCL; L3: the fourth ICS on the right midaxillary line (MAL); L4: the tenth ICS on the right MAL; L5: the second ICS on the left MCL; L6: the fifth ICS on the left MCL; L7: the fourth ICS on the left MAL; L8: the tenth ICS on the left MAL.

\paragraph{ICBHI 2017.}
There are two sources of data, the first source is from the Respiratory Research and Rehabilitation Laboratory of the School of Health Sciences, University of Aveiro. Sounds were recorded from the trachea and six chest locations: left and right front, back and side.
The second source is from Aristotle University of Thessaloniki. Sounds were sequentially collected  from six chest locations, as shown in Figure~\ref{subfig:icbhi2017}, with a digital stethoscope (WelchAllyn Meditron Master Elite Plus Stethoscope Model 5079-400 or 3M Litmann 3200).

\paragraph{Lung Sound.}
The data were collected by placing the stethoscope on various regions of the chest referred to as zones. Figure~\ref{subfig:lung_sound} illustrates the approximate boundaries of the chest zones. The chest zones are defined as follows:
1. Upper zone: located in the region under the clavicles and above the cardiac silhouette (i.e., the superior aspect of the hilum). Sometimes this region includes the apical zone, which is located above the inferior margin of clavicles.
2. Middle zone: located in the region between the superior and inferior aspect of the hilum.
3. Lower zone: located in the region enclosed by the inferior aspect of the hilum and the
hemidiaphragm.
Data from 10 chest zones were finally collected: anterior left upper, anterior right upper, anterior right middle, anterior right lower, posterior left lower, posterior left middle, posterior left upper, posterior right lower, posterior right middle and posterior right upper.

\paragraph{RespiratoryDatabase@TR.}
There are 16 channels in total, 4 channels for heart sounds focusing on the aorta, pulmonary artery, tricuspid valve, and mitral valve areas. 12 channels for lung sounds focusing on the upper lung, middle lung, lower lung, and costophrenic angle areas on the posterior and anterior sides of the chest.
As shown in figure~\ref{subfig:rd@tr}, the chest auscultation areas used in the physical examination include the posterior upper lung (L1-R1), posterior middle lung (L2-R2), posterior lower lung (L3-R3), posterior costophrenic angle lung (L5-R5), and anterior lower lung (L6-R6) on the left (L) and right (R) sides of the patient. The left image illustrates the chest auscultation areas on the anterior chest wall, while the middle image shows the posterior (back) chest auscultation areas.
Heart auscultation is performed at four distinct locations: the aorta (R7), pulmonary artery (L7), tricuspid valve (R8), and mitral valve (L8) areas. The cardiac auscultation areas are shown in the right figure of Figure~\ref{subfig:rd@tr}.

\paragraph{Circor 2022.}
Normal heart sounds are primarily generated by the vibrations of cardiac valves as they open and close during each cardiac cycle and the turbulence of blood into the arteries. The anatomical positions of heart valves relative to the chest wall determine the optimal auscultation position. Consequently, the stethoscope should be placed at the following positions for auscultation, as illustrated in Figure~\ref{subfig:circor2022}:
Aortic valve (1): second intercostal space, right sternal border; 
Pulmonary valve (2): second intercostal space, left sternal border; 
Tricuspid valve (3): left lower sternal border; 
Mitral valve (4): fifth intercostal space, midclavicular line (cardiac apex).

\paragraph{HSDReport.}
The auscultation protocol consists of recordings from five body sites: aortic region (right 2nd intercostal space), pulmonic region (left 2nd intercostal space, parasternal), Erb’s point (left 3rd intercostal space, also known as the left lower sternal border), tricuspid region (left 4th intercostal space, parasternal), mitral region (left 5th intercostal space, midclavicular).

\subsection{Sound Spectrogram}
\begin{figure}
    \centering
    \subfigure[Heart Sound Mel Spectrogram Examples]{
        \begin{minipage}[]{1\linewidth}
        \includegraphics[width=\linewidth]{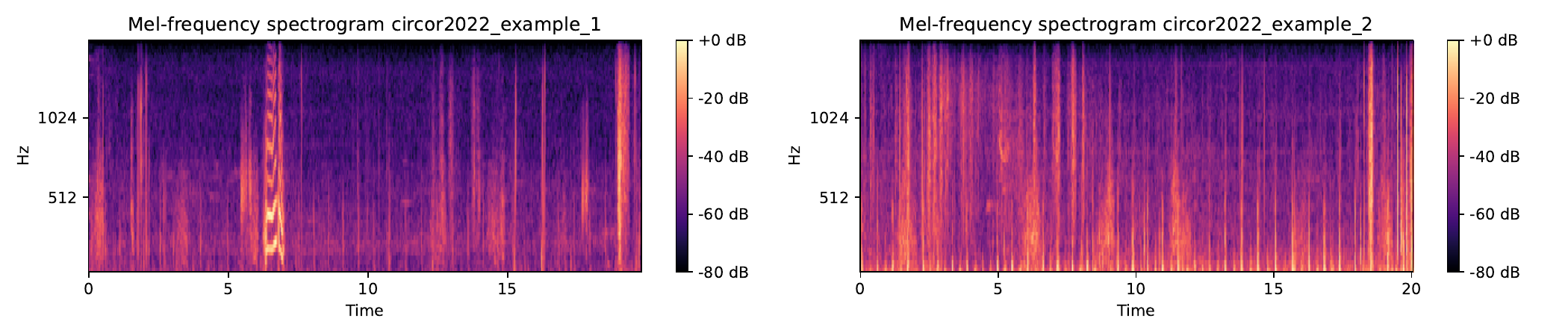}
        \end{minipage}
        \label{subfig:heart_sound_mel}
    }
    \subfigure[Lung Sound Mel Spectrogram Examples]{
        \begin{minipage}[]{1\linewidth}
        \includegraphics[width=\linewidth]{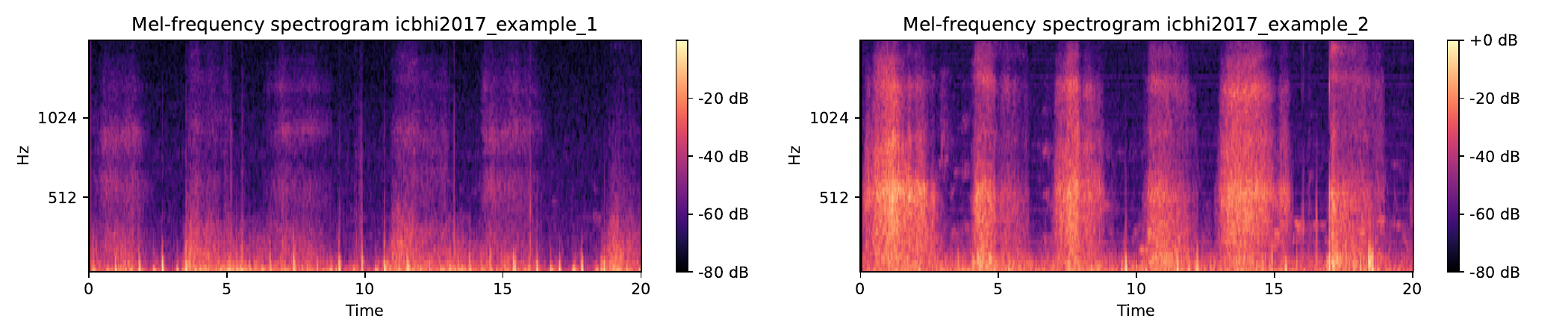}
        \end{minipage}
        \label{subfig:lung_sound_mel}
    }

    \subfigure[Bowel Sound Mel Spectrogram Examples]{
        \begin{minipage}[]{1\linewidth}
        \includegraphics[width=\linewidth]{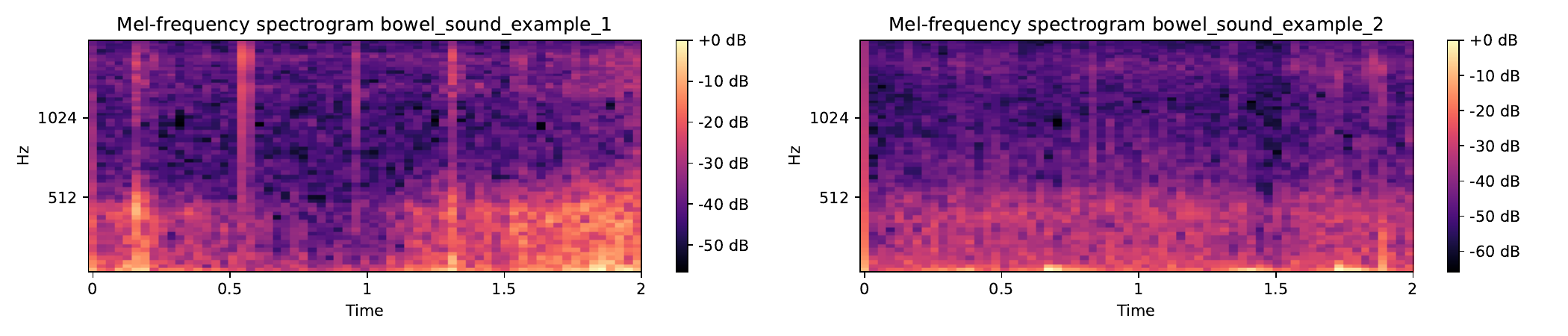}
        \end{minipage}
        \label{subfig:bowel_sound_mel}
    }

    \caption{Mel spectrograms of heart sounds, lung sounds, and bowel sounds, using audio from the Circor 2022, ICBHI 2017, and Bowel Sound datasets as examples, respectively.}
    \label{fig:sound_mel}
\end{figure}

As shown in Figure~\ref{fig:sound_mel}, two different audio samples were selected from the Circor 2022, Lung Sound, and Bowel Sound datasets. For each sample, 64-dimensional Mel spectrogram features were extracted to facilitate a visual comparison of the differences between the Mel features of these three types of sounds. Generally, bowel sound recordings are relatively short, averaging around 2 seconds, followed by lung sound recordings at approximately 20 seconds. In contrast, heart sound data can be significantly longer, exceeding 60 seconds. During the training phase, slice processing was applied to handle varying recording lengths.

\section{Downstream Task Description}
\label{apx:benchmark}

Here, we provide a detailed description of all 16 tasks defined in the benchmark. These tasks are organized into four categories:
\begin{itemize}
\item \textbf{Binary Classification (Tasks 3, 9, 14, 15, 16)}: Tasks that require predicting a binary outcome (normal/abnormal) from an audio recording.
\item \textbf{Multi-Class Classification (Tasks 1-2, 4-8, 10, 12-13)}: The tasks involve classifying an audio recording into one or multiple classes of several predefined categories.
\item \textbf{Multi-Label Classification (Task 11)}: The task involves classifying an audio recording into one of several predefined categories.
\end{itemize}
Below is a description of the downstream tasks performed using different datasets.

\paragraph{Task 1: SPRSound\_MC.}
Task 1 involves ternary classification at the record level, with the objective of categorizing respiratory sound records into three classes: Normal, Adventitious, and Poor Quality.

\paragraph{Task 2: SPRSound\_MC.}
Task 2 involves multi-class classification at the record level, with the objective of categorizing respiratory sound records into one of five classes: Normal, CAS, DAS, CAS \& DAS, or Poor Quality.

\paragraph{Task 3: HF\_Lung\_BC.}
Task 3 is a binary classification task aimed at classifying respiratory sound events as Normal or Abnormal.

\paragraph{Task 4: HF\_Lung\_MC.}
Task 4 is a four-class classification task aimed at classifying respiratory audio into four respiratory cycles: Inhalation, Exhalation, CAS, and DAS. A respiratory cycle refers to the sequence of events during which a person inhales (inspiration) and exhales (expiration) a given volume of air through the respiratory system.

\paragraph{Task 5: HF\_Lung\_MC.}
Task 5 is a six-class classification task. Building upon Task 4, the categories CAS and DAS are further subdivided to classify respiratory audio into the following six classes: Inhalation, Exhalation, Wheeze, Stridor, Rhonchi, and Crackle.

\paragraph{Task 6: ICBHI2017\_MC.}
Task 7 is a four-class lung sound classification task, aiming to classify respiratory events into four types of respiratory cycles: Normal, Crackle, Wheeze, and Crackle \& Wheeze. 

\paragraph{Task 7: LungSound\_MC.}
Task 6 is a seven-class classification task, aiming to classify audio into the following categories based on breathing cycles: Normal, Crepitation, Wheeze, Crackle, Bronchi, Wheeze \& Crackle, and Bronchi \& Crackle.

\paragraph{Task 8: Circor2022\_MC.}
Task 8 is a three-class classification task, aiming to classify heart murmurs into three categories: Present, Absent, and Unknown.

\paragraph{Task 9: BowelSound\_BC.}
Task 9 is a binary classification task aimed at assessing the occurrence of bowel sounds in the recordings.

\paragraph{Task 10: ICBHI2017\_MC\_Disease.}
Task 9 is a six-category classification task for lung diseases, where the goal is to predict the presence of six conditions based on the entire audio: Bronchiectasis, Bronchiolitis, COPD, Pneumonia, URTI, and Healthy.

\paragraph{Task 11: LungSound\_ML.}
Task 10 is a disease detection task involving eight categories, aiming to detect the presence of the following conditions based on audio: Normal, Asthma, Pneumonia, COPD, Bronchitis (BRON), Heart Failure, Lung Fibrosis, and Pleural Effusion.

\paragraph{Task 12: RD@TR\_MC.}
Task 11 is a five-class classification task that categorizes the data into COPD severity levels ranging from 0 (under risk) to 4 (very severe level) based on audio. For a more detailed description of the grading criteria, please refer to \cite{altan2017multimedia} and \cite{roisin2016chronic}. The goal of this task is to predict the severity of COPD using lung sounds.

\paragraph{Task 13: Korean\_MC.}
Task 12 is a five-class classification task that aims to detect the presence of heart diseases based on audio, including Normal (N), Aortic Stenosis (AS), Mitral Regurgitation (MR), Mitral Stenosis (MS), and Mitral valve prolapse (MVP).



\paragraph{Task 14: Cinc2016\_BC.}
Task 13 is a binary classification task that aims to classify sound events as Normal or Abnormal.

\paragraph{Task 15: Circor2022\_BC.}
Task 14 is a binary classification task that aims to classify sounds as normal or abnormal for detecting cardiac abnormalities.

\paragraph{Task 16: HSDReport\_BC.}
Task 15 is a binary classification task that aims to classify cardiac sound events as Normal or Abnormal.

\section{Pretraining Results}

\subsection{Pretraining Loss \& Accuracy}
\begin{figure}
    \centering
    \includegraphics[width=\linewidth]{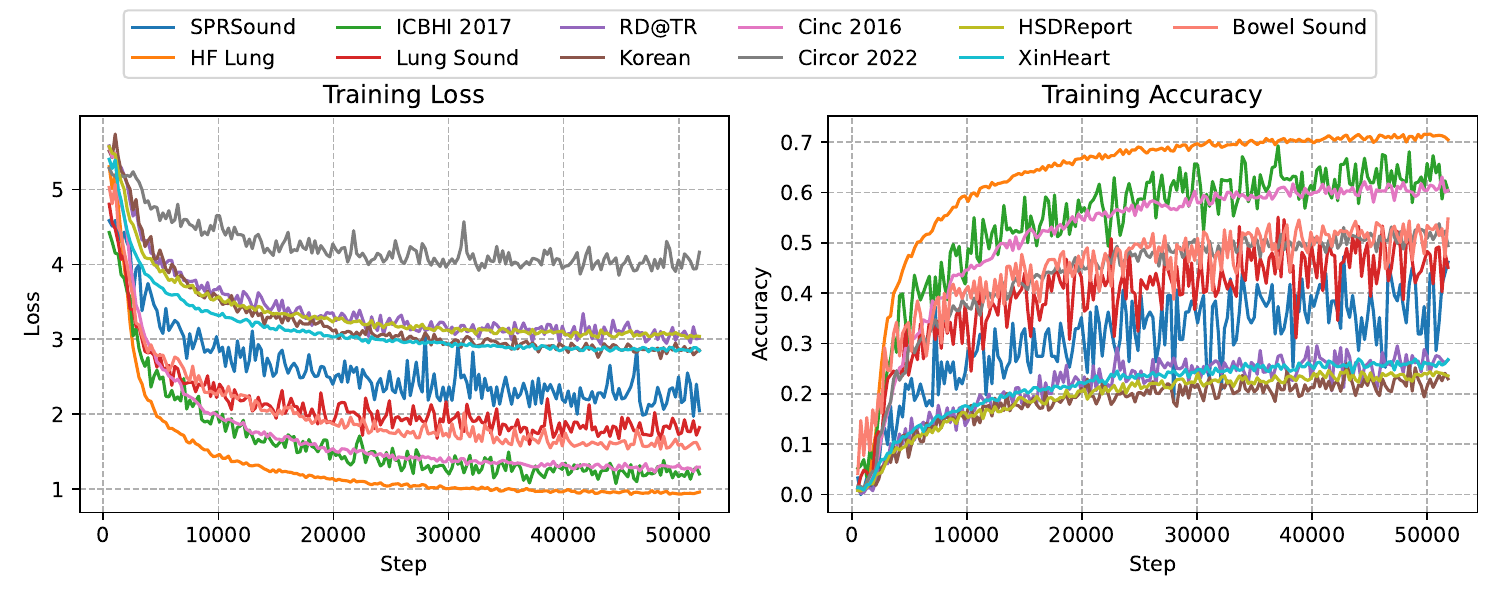}
    \caption{Training loss and contrastive instance discrimination accuracy of our model during the pretraining process.}
    \label{fig:train_curve}
\end{figure}

\begin{figure}
    \centering
    \includegraphics[width=\linewidth]{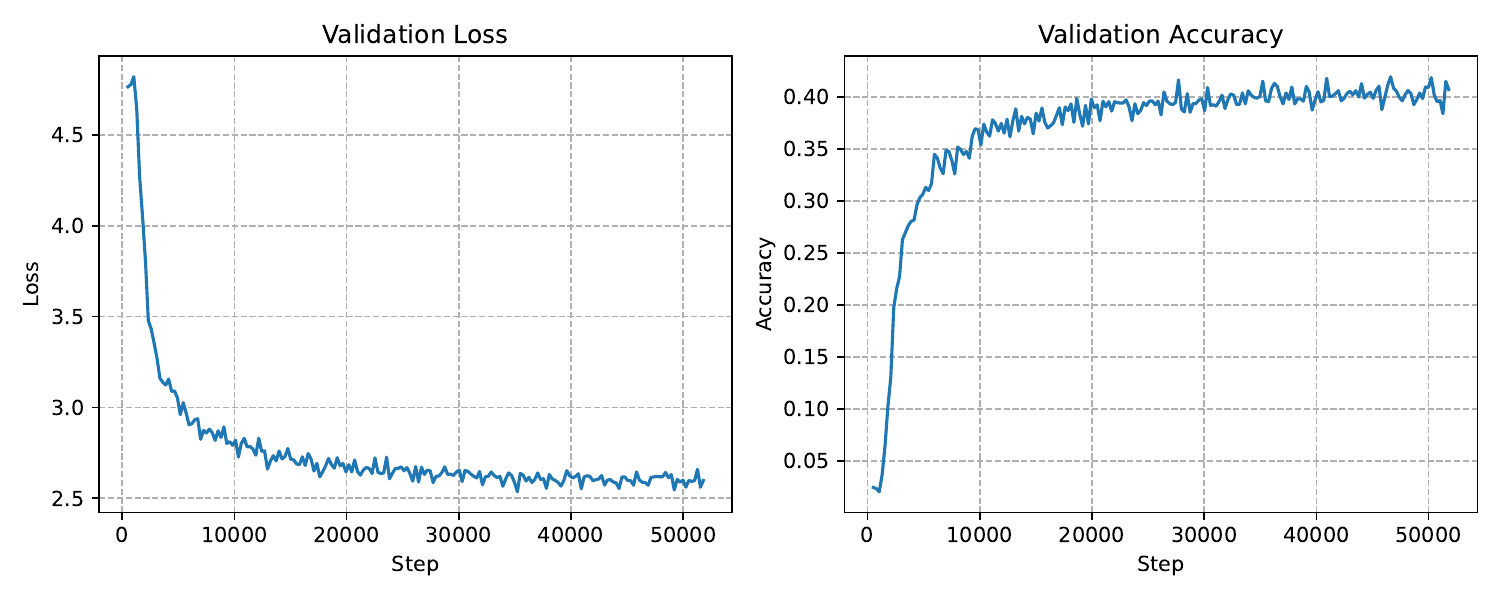}
    \caption{Validation loss and contrastive instance discrimination accuracy of our model during the pretraining process.}
    \label{fig:valid_curve}
\end{figure}

We exhibit the loss and accuracy during the pretraining process. As Figure~\ref{fig:train_curve} shows, the training loss of different subsets of the data converges at different speeds and levels, as well as the accuracy rate is gradually increasing. Similarly, Figure~\ref{fig:valid_curve} presents the evolution of the validation loss and accuracy on the combined subset of all the data resources, which also demonstrates a continued decay until convergence.

The varied convergence speed is relevant to the heterogeneity in data quality and modality. Meanwhile, the degree of task difficulty (e.g., the crop length and batch size) also affects the convergence level. In general, the longer the crop length is and the smaller the batch size is, the easier the discriminative task will be, which leads to lower training loss but worse model generalizability. 

\subsection{Embedding Distribution}

\begin{figure}
    \centering
    \subfigure[SPRSound]{
        \begin{minipage}[]{0.3\linewidth}
        \includegraphics[width=\linewidth]{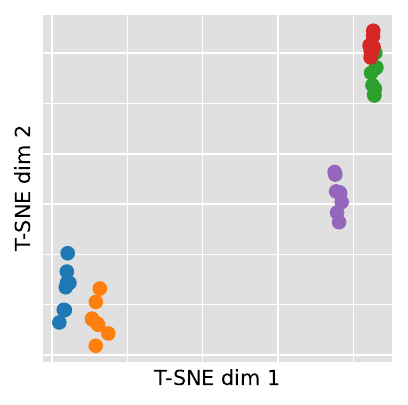}
        \end{minipage}
    }
    \subfigure[Lung Sound]{
        \begin{minipage}[]{0.3\linewidth}
        \includegraphics[width=\linewidth]{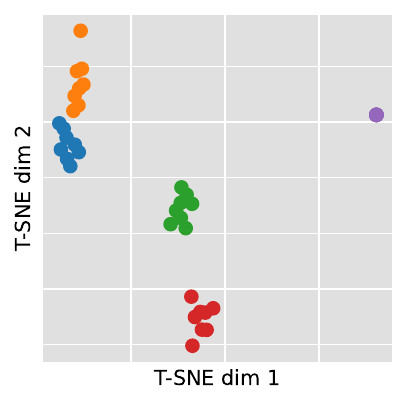}
        \end{minipage}
    }
    \subfigure[RD@TR]{
        \begin{minipage}[]{0.3\linewidth}
        \includegraphics[width=\linewidth]{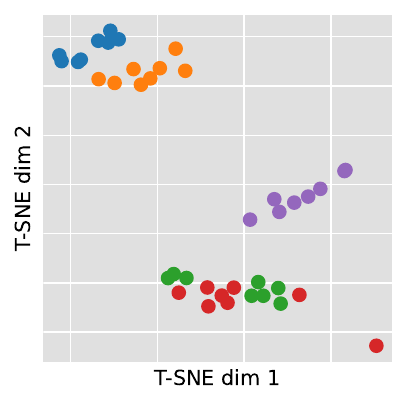}
        \end{minipage}
    }
    \subfigure[Korean]{
        \begin{minipage}[]{0.3\linewidth}
        \includegraphics[width=\linewidth]{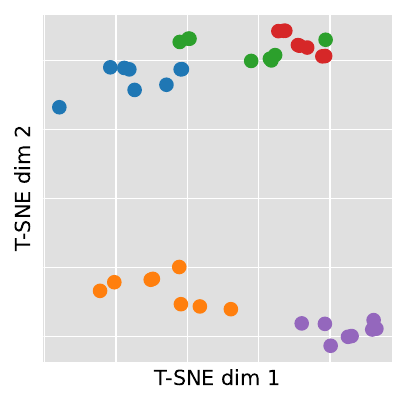}
        \end{minipage}
    }
    \subfigure[Cinc 2016]{
        \begin{minipage}[]{0.3\linewidth}
        \includegraphics[width=\linewidth]{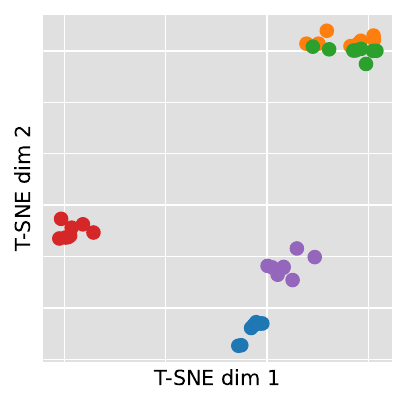}
        \end{minipage}
    }
    \subfigure[Circor 2022]{
        \begin{minipage}[]{0.3\linewidth}
        \includegraphics[width=\linewidth]{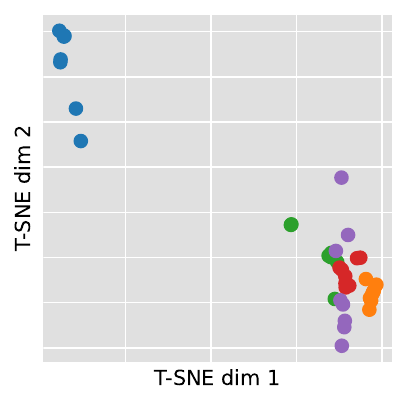}
        \end{minipage}
    }
    \subfigure[HSDReport]{
        \begin{minipage}[]{0.3\linewidth}
        \includegraphics[width=\linewidth]{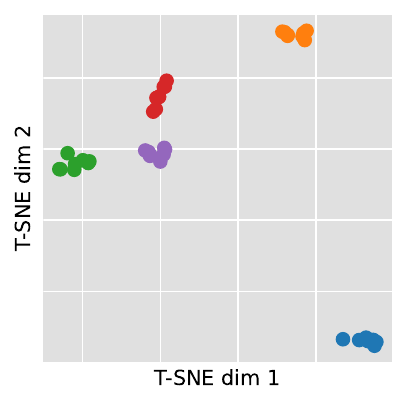}
        \end{minipage}
    }
    \subfigure[XHheartSound]{
        \begin{minipage}[]{0.3\linewidth}
        \includegraphics[width=\linewidth]{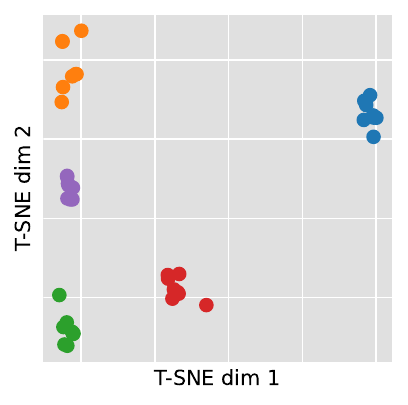}
        \end{minipage}
    }
    \subfigure[Bowel Sound]{
        \begin{minipage}[]{0.3\linewidth}
        \includegraphics[width=\linewidth]{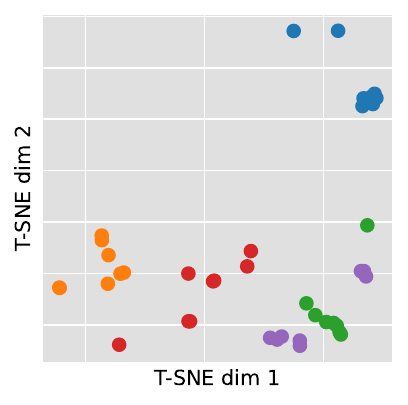}
        \end{minipage}
    }

    \caption{Biodimensional T-SNE visualizations results of the features of our model on the validation set of different datasets. Each dot represents a segment, and the dots with the same color denote that they are from the same recording.}
    \label{fig:tsne}
\end{figure}

To clarify the discriminative capability of the features extracted by our model, we visualize the T-SNE distribution as shown in Figure~\ref{fig:tsne}. Specifically, 8 segments are randomly cropped from each recording, and 5 samples for each dataset are visualized. As depicted in Figure~\ref{fig:tsne}, the segments from the same recording are close to each other while far away from others in the embedding space, which reveals that our model can capture the underlying homogeneous characteristics of the same recording despite the variance introduced by random cropping.

\section{Additional Evaluation Results}

\subsection{Detailed Class-wise Evaluation in Multi-class Scenarios}

\begin{figure}[t]
    \centering
    \includegraphics[width=\linewidth]{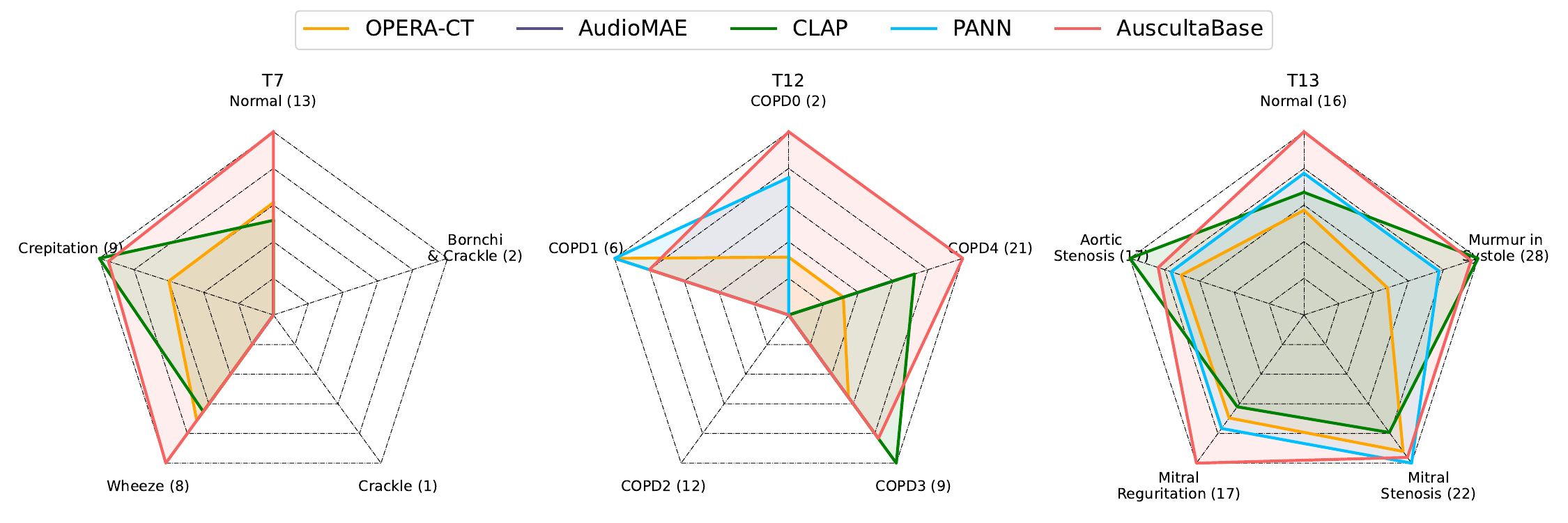}
    \caption{Class-wise F1 scores for multi-class classification tasks (T7, T12, T13). Each label at the corner of the radar chart represents a category, with the number in parentheses indicating the count of samples of that category within the test set. We exclude classes ``Bronchi" and ``Wheeze \& Crackle" for T7 because they are not represented in the test set.}
    \label{fig:classwise_f1}
\end{figure}

Figure~\ref{fig:classwise_f1} presents radar charts of class-wise F1 scores for multi-class classification tasks (T7, T12, and T13). Our model (highlighted in red) consistently demonstrates competitive or superior performance across both majority and minority classes. For instance, in T7—a task marked by a significant class imbalance—AuscultaBase outperforms competitors in minority categories such as “Crepitation” and “Wheeze”. Similarly, in T12, which involves grading the severity of COPD (ranging from COPD0 to COPD4), our model maintains high F1 scores even for the underrepresented classes. This balanced performance across classes underlines the model’s robustness in handling real-world datasets, where class imbalances are common and can lead to biased diagnostic outcomes.

\subsection{Enhanced Performance Under Full Fine-tuning}

We further investigate model performance under full fine-tuning conditions for both binary and multi-class tasks. For binary tasks (T14 and T16), Table~\ref{tab:full-finetune-BC} shows that AuscultaBase achieves significantly improved AUC scores after full fine-tuning, with task T14 reaching an AUC of 0.978. Such improvements are particularly valuable in clinical scenarios that demand high sensitivity and specificity. For the multi-class task T12—despite its relatively small training sample size (454 samples)—AuscultaBase achieved a Macro-F1 of 41.78 (linear) and 55.14 (full fine-tuning), with corresponding Micro-F1 scores of 49.00 and 57.00, respectively (Table~\ref{tab:full-finetune-MC}). These gains underscore the model’s adaptability and its capacity to capture complex inter-class relationships, even when data are limited—a common challenge in medical diagnostics.

\begin{table}[t]
    \centering
    \resizebox{\linewidth}{!}{
    \begin{tabular}{c|c|cccccc}
        \toprule
        \textbf{ID} & \textbf{\#} & \textbf{Method} & \textbf{OPERA-CT} & \textbf{AudioMAE} & \textbf{CLAP} & \textbf{PANN} & \textbf{AuscultaBase} \\
        \midrule
        \multirow{2}{*}{T14} & \multirow{2}{*}{2918} & Linear & 0.850$\pm$0.003 & 0.872$\pm$0.003 & 0.835$\pm$0.023 & 0.824$\pm$0.001 & \textbf{0.920}$\pm$0.008 \\
         & & Full & 0.961$\pm$0.007 & 0.951$\pm$0.007 & 0.857$\pm$0.003 & 0.959$\pm$0.006 & \textbf{0.978}$\pm$0.003 \\
        \midrule
        \multirow{2}{*}{T16} & \multirow{2}{*}{3294} & Linear & 0.795$\pm$0.005 &	0.691$\pm$0.022 &	0.766$\pm$0.010 &	0.484$\pm$0.020 &	\textbf{0.802}$\pm$0.008 \\
         & & Full & 0.784$\pm$0.001 & 0.760$\pm$0.063 & 0.776$\pm$0.004 & 0.652$\pm$0.037 & \textbf{0.824}$\pm$0.013 \\
        \bottomrule
    \end{tabular}
    }
    \caption{AUC for linear probing and full fine-tuning on T14 and T16. \# represents the number of training samples for the task. AUC ranges from 0 to 1, and \textcolor{red}{higher} is better.}
    
    \label{tab:full-finetune-BC}
\end{table}

\begin{table}[t]
    \centering
    \resizebox{\linewidth}{!}{
    \begin{tabular}{c|c|l|cccccc}
        \toprule
        \textbf{ID} & \textbf{\#} & \textbf{Metric} & \textbf{Method} & \textbf{OPERA-CT} & \textbf{AudioMAE} & \textbf{CLAP} & \textbf{PANN} & \textbf{AuscultaBase} \\
        \midrule
        \multirow{4}{*}{T12} & \multirow{4}{*}{454} & \multirow{2}{*}{Macro-F1} & Linear & 30.74$\pm$0.77 & 17.00$\pm$5.17 & 27.68$\pm$2.07 & 30.15$\pm$5.57 & \textbf{41.78}$\pm$1.83 \\
        & & & Full & 45.04$\pm$2.97 &	48.48$\pm$0.97 &	32.87$\pm$7.27 &	53.01$\pm$8.63 &	\textbf{55.14}$\pm$4.01 \\
        \cmidrule{3-9}
        & & \multirow{2}{*}{Micro-F1} & Linear & 36.00$\pm$2.00 & 43.00$\pm$1.00 & 41.00$\pm$3.00 & 40.00$\pm$4.00 & \textbf{49.00}$\pm$1.00 \\
        & & & Full & 49.00$\pm$3.00 &	54.00$\pm$3.00 &	47.00$\pm$1.00 &	54.00$\pm$8.00 &	\textbf{57.00}$\pm$5.00 \\
        \bottomrule
    \end{tabular}
    }
    \caption{Macro-F1 and Micro-F1 scores for linear probing and full finetuning on T12. Macro-F1 and Micro-F1 range from 0 to 100, and \textcolor{red}{higher} is better.}
    
    \label{tab:full-finetune-MC}
\end{table}




\end{document}